\documentclass[letter,11pt]{article}
 \pdfoutput=1
 \usepackage{jheppub}
 
 \usepackage{graphicx}
 \usepackage{hyperref}
 \usepackage[utf8]{inputenc}
 \usepackage{amssymb}
\usepackage{multirow}
\usepackage{soul}
 \usepackage{ulem}
 
 \usepackage{booktabs}
\usepackage{mathrsfs}
\usepackage{epsfig}
\usepackage{graphicx}
\usepackage{subfigure}
\usepackage{dcolumn}
\usepackage{bm}
\usepackage{amsmath}
\usepackage{slashed}
\usepackage{multirow}
\usepackage{color}
\usepackage{dcolumn}
\usepackage{bm}
\usepackage{color}
\usepackage[dvipsnames]{xcolor}

\newcommand{\ord}{{\cal O}}
\def\beq{\begin{equation}}
\def\eeq{\end{equation}}
\def\eeqn{\end{equation}}
\newcommand\iden{\leavevmode\hbox{\small1\normalsize\kern-.33em1}}

\newcommand{\calL}{{\cal L}}

\newcommand{\bea} {\begin{eqnarray}}
\newcommand{\eea} {\end{eqnarray}}

\def\tb {t_\beta}
\def\sb  {s_{\beta}}
\def\cb  {c_{\beta}}

\def\lam{\lambda}

\def\hm{{\hat m}}

\let\jnfont=\rm
\def\NPB#1 {{\jnfont Nucl.\ Phys.\ B }{\bf #1} }
\def\PLB#1 {{\jnfont Phys.\ Lett.\ B }{\bf #1} }
\def\EPJC#1 {{\jnfont Eur.\ Phys.\ Jour.\ C }{\bf #1} }
\def\PRD#1 {{\jnfont Phys.\ Rev.\ D }{\bf #1} }
\def\PRL#1 {{\jnfont Phys.\ Rev.\ Lett.\ }{\bf #1} }
\def\MPLA#1 {{\jnfont Mod.\ Phys.\ Lett.\ A }{\bf #1} }
\def\JPG#1 {{\jnfont J.\ Phys.\ G }{\bf #1} }
\def\CTP#1 {{\jnfont Commun.\ Theor.\ Phys.\ }{\bf #1} }
\def\JHEP#1 {{\jnfont JHEP \ }{\bf #1} }
\def\NPPS#1 {{\jnfont Nucl.\ Phys.\ Proc.\ Suppl.\ }{\bf #1} }
\def\CPC#1 {{\jnfont Comput.\ Phys.\ Commun.\ }{\bf #1} }
\def\CPL#1 {{\jnfont Chin.\ Phys.\ Lett. }{\bf #1} }
\def\APPB#1 {{\jnfont Acta\ Phys.\ Polon.\ B }{\bf #1} }

\def\lsim{\raise0.3ex\hbox{$<$\kern-0.75em\raise-1.1ex\hbox{$\sim$}}}
\def\gsim{\raise0.3ex\hbox{$>$\kern-0.75em\raise-1.1ex\hbox{$\sim$}}}
\def\PR#1 {{\jnfont Phys.\ Rept. }{\bf #1} }
\def\CHC#1 {{\jnfont Chin.\ Phys.\ C }{\bf #1} }
\def\NIMA#1 {{\jnfont Nucl.\ Instrum.\ Meth.\ A }{\bf #1} }
\def\JCAP#1 {{\jnfont JCAP \ }{\bf #1} }
\def\ASA#1 {{\jnfont Astron.\ Astrophys.\ A }{\bf #1} }  

\begin{document}

\title{\ \\[10mm] Inflation, electroweak phase transition, and Higgs searches at the LHC in the two-Higgs-doublet model}

\author{Lei Wang}
 \affiliation{Department of Physics, Yantai University, Yantai
264005, P. R. China}


\abstract{Combining the Higgs searches at the LHC, 
we study the Higgs inflation in the type-I and type-II two-Higgs-doublet models with non-minimally couplings to gravity.
After imposing relevant theoretical and experimental constraints, we find that the Higgs inflation imposes stringent constraints on
the mass splitting between $A$, $H^\pm$, and $H$, and they tend to be nearly degenerate in mass with increasing of their masses.
The direct searches for Higgs at the LHC can exclude many points achieving Higgs inflation in the region of $m_H~(m_A)<$ 450 GeV in the type-I model,
and impose a lower bound on $\tan\beta$ for the type-II model. The Higgs inflation disfavors the wrong sign Yukawa coupling region of 
type-II model. In the parameter space achieving the Higgs inflation, the type-I and type-II models can produce a first order electroweak phase transition,
but $v_c/T_c$ is much smaller than 1.0.}

\maketitle

\section{Introduction}
The cosmic inflation during very early phase of the Universe can explain a number of cosmological problems, such as the horizon and flatness problems \cite{infla1,infla2,infla3}. A attractive scenario is the Standard Model (SM) Higgs as the inflaton field, which is non-minimally coupled to 
gravity \cite{non-mini-grav,higgs-infla1,higgs-infla2}. The SM Higgs plays an important role in particle physics, and its properties 
can be measured at the LHC. However, the current LHC data set the SM Higgs mass, $m_h \approx 125$ GeV \cite{atlas125,cms125}, which hints that
 the SM Higgs self-coupling runs to be negative values well below the Planck scale
or the inflationary scale \cite{smrge-1,smrge-2,smrge-3}. The SM vacuum becomes unstable before the non-minimal coupling becomes dominant. 
Therefore, the Higgs sectors of the SM need to be extended to achieve the Higgs inflation.

The two-Higgs-doublet model (2HDM) is a simple extension of SM by introducing a second $SU(2)_L$
Higgs doublet, which contains two neutral CP-even Higgs bosons $h$ and $H$, one neutral pseudoscalar $A$, and 
a pair of charged Higgs $H^\pm$ \cite{2hdm}. The ATLAS and CMS experimental data show that the properties of the discovered 125 GeV boson 
are well consistent with the SM Higgs boson. In addition, no excesses
are observed in the searches for the additional Higgs. Therefore, the searches for Higgs at the LHC
can impose stringent constraints on new physics models, especially for the 2HDM. 
In this paper, we consider the recent LHC Higgs data, and discuss the Higgs inflation in the type-I 2HDM \cite{i-1,i-2}
and type-II 2HDM \cite{i-1,ii-2}. There have been some studies on the inflation in the inert doublet model \cite{infla-inert1,infla-inert2}
and general 2HDM \cite{infla-g2hdm}. Next, we will combine the inflation to discuss the electroweak phase transition (EWPT) 
in the early universe, and the electroweak baryogenesis mechanism requires a strongly first order EWPT (FOEWPT) to give
 a successful explanation of the observed baryon asymmetry of the universe (BAU)~\cite{Sakharov:1967dj}. 
The EWPT in the 2HDM has been extensively studied in the Refs. \cite{PT_2HDM1-1,PT_2HDM1-2, PT_2HDM1-3, PT_2HDM1-4,
PT_2HDM1-5,PT_2HDM1-6,PT_2HDM1-7,PT_2HDM1-8,PT_2HDM1-9,PT_2HDM1.5, PT_2HDM2, 1711.09849, PT_2HDM3,pt2h-lwang,pt2h-wsu,pt2h-zhh}.

The paper is organized as follows. In Sec. II we will introduce the type-I and type-II 2HDMs with non-minimal couplings to gravity
 and inflation dynamics. In Sec. III we show the parameter space achieving the Higgs inflation after imposing relevant theoretical and experimental constraints.
In Sec. IV, we combine inflation to discuss the EWPT.
Finally, we give our conclusion in Sec. V.

\section{Two-Higgs-doublet model with non-minimally couplings to gravity}
\subsection{Two-Higgs-doublet model}
In the type-I and type-II 2HDMs, the Higgs potential with a soft $Z_2$ symmetry breaking
can be written as \cite{2h-poten}
\begin{eqnarray} \label{V2HDM} \mathcal{V}_{tree} &=& m_{11}^2
(\Phi_1^{\dagger} \Phi_1) + m_{22}^2 (\Phi_2^{\dagger}
\Phi_2) - \left[m_{12}^2 (\Phi_1^{\dagger} \Phi_2 + \rm h.c.)\right]\nonumber \\
&&+ \frac{\lambda_1}{2}  (\Phi_1^{\dagger} \Phi_1)^2 +
\frac{\lambda_2}{2} (\Phi_2^{\dagger} \Phi_2)^2 + \lambda_3
(\Phi_1^{\dagger} \Phi_1)(\Phi_2^{\dagger} \Phi_2) + \lambda_4
(\Phi_1^{\dagger}
\Phi_2)(\Phi_2^{\dagger} \Phi_1) \nonumber \\
&&+ \left[\frac{\lambda_5}{2} (\Phi_1^{\dagger} \Phi_2)^2 + \rm
h.c.\right].
\end{eqnarray}
We consider a CP-conserving case in which all $\lambda_i$ and
$m_{12}^2$ are real. The two complex Higgs doublet fields $\Phi_1$ and $\Phi_2$ have hypercharge $Y = 1$ and are expanded as
\begin{equation}
\Phi_1=\left(\begin{array}{c} \phi_1^+ \\
\frac{1}{\sqrt{2}}\,(v_1+\phi_1+ia_1)
\end{array}\right)\,, \ \ \
\Phi_2=\left(\begin{array}{c} \phi_2^+ \\
\frac{1}{\sqrt{2}}\,(v_2+\phi_2+ia_2)
\end{array}\right),
\end{equation}
with $v_1$ and $v_2$ being the electroweak vacuum expectation values (VEVs) and $v^2 = v^2_1 + v^2_2 = (246~\rm GeV)^2$. 
We define the ratio of the two VEVs as $\tan\beta=v_2 /v_1$. 
 After spontaneous electroweak symmetry breaking, the mass eigenstates are obtained from the original fields by the rotation matrices,
\begin{eqnarray}
\left(\begin{array}{c}H \\ h \end{array}\right) =  \left(\begin{array}{cc}\cos\alpha & \sin\alpha \\ -\sin\alpha & \cos\alpha \end{array}\right)  \left(\begin{array}{c} \phi_1 \\ \phi_2 \end{array}\right) , \\
\left(\begin{array}{c}G^0 \\ A \end{array}\right) =  \left(\begin{array}{cc}\cos\beta & \sin\beta \\ -\sin\beta & \cos\beta \end{array}\right)  \left(\begin{array}{c} a_1 \\ a_2 \end{array}\right) , \\
\left(\begin{array}{c}G^{\pm} \\ H^{\pm} \end{array}\right) =  \left(\begin{array}{cc}\cos\beta & \sin\beta \\ -\sin\beta & \cos\beta \end{array}\right)  \left(\begin{array}{c} \phi^{\pm}_1 \\ \phi^{\pm}_2 \end{array}\right).
\end{eqnarray}
The $G^0$ and $G^\pm$ are Goldstones which are eaten by gauge bosons $Z$ and $W^\pm$. 
The remaining physical states are two neutral
CP-even states $h$, $H$, one neutral pseudoscalar $A$, and a pair of charged
scalars $H^{\pm}$.

The type-I and type-II 2HDMs have different $Z_2$ parity assignments for the right field of fermion.
The Yukawa interactions of type-I model are 
 \bea
- {\cal L} &=&Y_{u2}\,\overline{Q}_L \, \tilde{{ \Phi}}_2 \,u_R
+\,Y_{d2}\,
\overline{Q}_L\,{\Phi}_2 \, d_R\, + \, Y_{\ell 2}\,\overline{L}_L \, {\Phi}_2\,e_R+\, \mbox{h.c.}\,. \eea
The Yukawa interactions of type-II 2HDM are
 \bea
- {\cal L} &=&Y_{u2}\,\overline{Q}_L \, \tilde{{ \Phi}}_2 \,u_R
+\,Y_{d1}\,
\overline{Q}_L\,{\Phi}_1 \, d_R\, + \, Y_{\ell 1}\,\overline{L}_L \, {\Phi}_1\,e_R+\, \mbox{h.c.}\,. \eea 
where
$Q_L^T=(u_L\,,d_L)$, $L_L^T=(\nu_L\,,l_L)$,
$\widetilde\Phi_{1,2}=i\tau_2 \Phi_{1,2}^*$, and $Y_{u2}$,
$Y_{d1,d2}$ and $Y_{\ell 1,\ell 2}$ are $3 \times 3$ matrices in family
space.

The Yukawa couplings of the neutral Higgs bosons with respect to the SM are 
\bea\label{hffcoupling} &&
y_{h}^{f_i}=\left[\sin(\beta-\alpha)+\cos(\beta-\alpha)\kappa_f\right], \nonumber\\
&&y_{H}^{f_i}=\left[\cos(\beta-\alpha)-\sin(\beta-\alpha)\kappa_f\right], \nonumber\\
&&y_{A}^{f_i}=-i\kappa_f~{\rm (for~u)},~~~~y_{A}^{f_i}=i \kappa_f~{\rm (for~d,~\ell)},\nonumber\\ 
&&{\rm with}~\kappa_d=\kappa_\ell=\kappa_u\equiv 1/\tan\beta  {\rm~ for~ type-I},\nonumber\\
&&{\rm~~~~~~}~\kappa_u\equiv 1/\tan\beta {\rm~and~} \kappa_d=\kappa_\ell\equiv-\tan\beta  {\rm~ for~ type-II}.\eea 
The Yukawa interactions of the charged Higgs are 
\begin{align} \label{eq:Yukawa2}
 \mathcal{L}_Y & = - \frac{\sqrt{2}}{v}\, H^+\, \Big\{\bar{u}_i \left[\kappa_d\,(V_{CKM})_{ij}~ m_{dj} P_R
 - \kappa_u\,m_{ui}~ (V_{CKM})_{ij} ~P_L\right] d_j + \kappa_\ell\,\bar{\nu} m_\ell P_R \ell
 \Big\}+h.c.,
 \end{align}
where $i,j=1,2,3$. The neutral Higgs boson couplings with the gauge bosons normalized to the
SM are 
\beq
y^{V}_h=\sin(\beta-\alpha),~~~
y^{V}_H=\cos(\beta-\alpha),\label{hvvcoupling}
\eeq
where $V=Z,~W$.
In the type-II model, the 125 GeV Higgs $h$ is allowed to
have the SM-like coupling and wrong sign Yukawa coupling, 
\bea
&&y_h^{f_i}~\times~y^{V}_h > 0~{\rm for~SM-like~coupling},~~~\nonumber\\
&&y_h^{f_i}~\times~y^{V}_h < 0~{\rm for~wrong~sign~Yukawa~coupling}.\label{wrongsign}
\eea

\subsection{Inflation dynamics}
To examine the inflation dynamics, we give the relevant Jordan frame Lagrangian,
\begin{equation}
\frac{\calL_J}{\sqrt{-g}} = \frac{R}{2}+\Big(\xi_1|\Phi_1|^2+\xi_2|\Phi_2|^2\Big)R
-\left|D_\mu\Phi_1\right|^2-\left|D_\mu\Phi_2\right|^2-V\left(\Phi_1,\Phi_2\right) \, ,
\end{equation}
where we have set $m_{Pl}$ = 1. $R$ is the Ricci scalar, and $\xi_1$ and $\xi_2$ are dimensionless couplings of the doublet fields to gravity.

We make the conformal transformation on the metric,
\beq
g^E_{\mu\nu}=g_{\mu\nu}\Omega^2 {~\rm with~}
\Omega^2\equiv 1+2\xi_1|\Phi_1|^2+2\xi_2|\Phi_2|^2 \, ,
\eeq
and obtain the Einstein frame action without the gauge interactions \cite{infla-inert1,1003.1159}
\begin{align}
\label{einsteinaction}
\frac{\calL_E}{\sqrt{-g_E}} = & \frac{R_E}{2}-\frac{3}{4}\Big[\partial_\mu \log \left(1+2\xi_1|\Phi_1|^2+2\xi_2|\Phi_2|^2\right)\Big]^2
-\frac{|\partial_\mu\Phi_1|^2+|\partial_\mu\Phi_2|^2}{1+2\xi_1|\Phi_1|^2+2\xi_2|\Phi_2|^2} \nonumber\\
&- V_E(\Phi_1,\Phi_2) \, \\
& V_E(\Phi_1,\Phi_2) =  \frac{V}{\left(1+2\xi_1|\Phi_1|^2+2\xi_2|\Phi_2|^2\right)^2} \, .
\end{align}

To examine inflation dynamics, we take two Higgs doublets as
\begin{equation}
\Phi_1=\frac{1}{\sqrt{2}}\left(\begin{array}{l} 0 \\ h_1 \end{array} \right) \, , \quad
\Phi_2=\frac{1}{\sqrt{2}}\left(\begin{array}{l} 0 \\ h_2 e^{i\theta} \end{array} \right) \, .  \label{higgsol}
\end{equation}
Ignoring the mass terms, the Einstein action in terms of field $\phi^i = {h_1, h_2, \theta}$ becomes
\begin{equation}\label{eq-po0}
\frac{{\cal L}_E}{\sqrt{-g_E}}= \frac{R_E}{2}-\frac{1}{2}S_{ij}\partial_\mu \phi^i\partial^\mu\phi^j-V_E(\phi^i) \, ,
\end{equation}
where
\begin{align}\label{kine-h1h2}
S_{ij} = & \frac{1}{1+\xi_1h^2_1+\xi_2 h^2_2}
\begin{pmatrix}
1+\dfrac{6\xi^2_1 h^2_1}{1+\xi_1h^2_1+\xi_2 h^2_2} & \dfrac{6\xi_1\xi_2h_1h_2}{1+\xi_1h^2_1+\xi_2 h^2_2} & 0
\\
\dfrac{6\xi_1\xi_2h_1h_2}{1+\xi_1h^2_1+\xi_2 h^2_2} & 1+\dfrac{6\xi^2_2 h^2_2}{1+\xi_1h^2_1+\xi_2 h^2_2}  & 0
\\
0 & 0 & h^2_2
\end{pmatrix} \, ,
\\
V_E(\phi^i) = & \frac{\lambda_1h^4_1+\lambda_2 h^4_2+2(\lambda_3+\lambda_4)h^2_1h^2_2+2\lambda_5 h^2_1h^2_2\cos(2\theta)}{8\left(1+\xi_1h^2_1+\xi_2 h^2_2 \right)^2} \, .
\label{epot2}
\end{align}
We redefine the scalar fields as follows \cite{1105.2284},
\begin{align}
\varphi= & \sqrt{\frac{3}{2}}\log(1+\xi_1 h^2_1+\xi_2 h^2_2) \, ,  \label{newfield1}
\\
\rho = & \frac{h_2}{h_1} \, ,  \label{newfield2}
\end{align}
and obtain the potential,
\begin{align}
\label{potentialwomin}
V_E(\varphi,\rho,\theta)= & \frac{\lambda_1+\lambda_2 \rho^4+2(\lambda_3+\lambda_4) \rho^2+2\lambda_5 \rho^2\cos(2\theta)}{8\left(\xi_1+\xi_2 \rho^2\right)^2}\,\left(1-e^{-2\varphi/\sqrt{6}}\right)^2  \, .
\end{align}
For $\lambda_5 \ll \lambda_1, \lambda_2, \lambda_3, \lambda_4$, the $\theta$ term of the potential in Eq. (\ref{potentialwomin}) 
does not affect the stabilization of the orthogonal
mode $\rho$. If the parameters satisfy the following conditions, \beq
\lambda_2\xi_1-(\lambda_3+\lambda_4)\xi_2 >  0, ~~~  \lambda_1\xi_2-(\lambda_3+\lambda_4)\xi_1 > 0,
\eeq
$\rho$ is stabilized at a finite value 
$\rho_0^2=\frac{\lambda_1\xi_2-(\lambda_3+\lambda_4)\xi_1}{\lambda_2\xi_1-(\lambda_3+\lambda_4)\xi_2}$ by the requirement of the
 potential extrema. Thus, we obtain the $\rho$ independent part of potential,
\begin{equation}\label{vrho}
V_{\rho\text{-indep}} = \frac{\lambda_{eff}}{8\left(\xi_1+\xi_2 \rho_0^2\right)^2} \,
\left(1-e^{-2\varphi/\sqrt{6}}\right)^2 \left(1 + \frac{2\lambda_5\rho_0^2}{\lambda_{eff}} \cos(2\theta)\right) \, ,
\end{equation}
where $\lambda_{eff}=\lambda_1+\lambda_2\rho_0^4 +2({\lambda_3+\lambda_4})\rho_0^2$. 
When $\mid\lambda_5\mid$ is smaller than $10^{-7}$, both the CP-even Higgs $\varphi$ and 
the pseudoscalar Higgs $\theta$ can drive the inflation, and the theoretical values of the inflationary observables 
are well consistent with the experimental values of Planck collaboration. The detailed discussions can be found in \cite{infla-inert1}. 
However, in the type-I and type-II models there is no symmetry to produce such small $\mid\lambda_5\mid$, and therefore 
we assume $\mid\lambda_5\mid \gg 10^{-7}$. As a result, the $\theta$ field does not drive the inflation but rather it is stabilized. 
After stabilizing $\theta$ at the minimum, we obtain the $\theta$ independent part of potential,
\begin{equation}\label{vrphi}
V_{\theta\text{-indep}} \approx \frac{\lambda_1+\lambda_2\rho^4 +2{\lambda}_L \rho^2}{8\left(\xi_1+\xi_2 \rho^2\right)^2} \,\left(1-e^{-2\varphi/\sqrt{6}}\right)^2  \, ,
\end{equation}
with ${\lambda}_L\equiv\lambda_3+\lambda_4-|\lambda_5|$. 
The $\varphi$ field can drive inflation, and the $\rho$ field needs to be stalized by the extremum condition of the potential of Eq. (\ref{vrphi}),
\beq
0=\frac{\partial V_{\theta\text{-indep}}}{\partial \rho }
= \frac{\rho\left(x_1 \rho^2-x_2\right)}{2\left(\xi_1+\xi_2 \rho^2\right)^3} \,\left(1-e^{-2\varphi/\sqrt{6}}\right)^2\, ,
\eeq
with $x_1\equiv \lambda_2\xi_1-\lambda_L\xi_2$ and $x_2\equiv \lambda_1\xi_2-\lambda_L\xi_1$. Thefore, there are three extrema at $\rho^2=0,~\infty$, and $x_2/x_1$.
 In addition to $\lambda_1>0$ and $\lambda_2>0$, the potential stability requires
  $\xi_1$ and $\xi_2$ to be positive at large field values. The double derivative of the potential is
\begin{equation}
\frac{\partial^2V_{\theta\text{-indep}}}{\partial \rho^2}
= \frac{\left(3x_1 \rho^2-x_2\right)\left(\xi_1+\xi_2 \rho^2\right) - 6\xi_2 \rho^2\left(x_1\rho^2-x_2\right)}{2\left(\xi_1+\xi_2 \rho^2\right)^4} \,\left(1-e^{-2\varphi/\sqrt{6}}\right)^2\, .
\end{equation}

\begin{itemize}
\item[(1)] $x_1 >0$ and $x_2 <  0$. The potential of Eq. (\ref{vrphi}) has one minimum only at $\rho^2=0$. 
For such case, the $h_1$ field plays the role of the inflation, which is called $h_1$-inflation. The potential of Eq. (\ref{vrphi}) becomes
\beq
V = \frac{\lambda_1}{8\xi^2_1} \left(1-e^{-2\varphi/\sqrt{6}}\right)^2. \label{eqslh1v}
\eeq

\item[(2)] $x_1 <0$ and $x_2 >  0$. The potential of Eq. (\ref{vrphi}) has one minimum only at $\rho^2=\infty$. 
As a result, the $h_2$ field plays the role of the inflation, which is called $h_2$-inflation. The potential of Eq. (\ref{vrphi}) becomes
\beq
V = \frac{\lambda_2}{8\xi^2_2} \left(1-e^{-2\varphi/\sqrt{6}}\right)^2. \label{eqslh2v}
\eeq

\item[(3)] $x_1 < 0$ and $x_2 <  0$.  The potential of Eq. (\ref{vrphi}) has two minima at $\rho^2=0$ and $\rho^2=\infty$. 
Both of the $h_1$-inflation and the $h_2$-inflation are feasible, and which one is chosen depends on the initial conditions for the $\rho$.

\item[(4)] $x_1 >0$ and $x_2 >  0$.  The potential of Eq. (\ref{vrphi}) has one minimum only at 
$\rho^2=\frac{x_2}{x_1}$. As a result, the inflaton is a mixture of $h_1$ and $h_2$.
\end{itemize}

In this paper we focus on the pure Higgs inflation, namely $h_1$-inflation and $h_2$-inflation.
For the $h_1$-inflation and $h_2$-inflation, the slow roll potentials of Eq. (\ref{eqslh1v}) and Eq. (\ref{eqslh2v}) do not depend on $\xi_2$ and $\xi_1$, respectively, and therefore we
simply take $\xi_2=0$ for the $h_1$-inflation and $\xi_1=0$ for the $h_2$-inflation.

\begin{itemize}
\item[(1)] $h_1$-inflation at $\rho^2=0$. Taking $\xi_2=0$ and replacing ($h_1$, $h_2$) with ($\rho$, $\varphi$), we can obtain the kinetic term from Eq. (\ref{eq-po0}) and Eq. (\ref{kine-h1h2}), 
\begin{align}
\label{h1-kin}
\calL_{kin-h_1} =&- \frac{1}{2}\left(1+\frac{\rho^2+1}{6\xi_1}\right)(\partial_\mu \varphi)^2-\frac{\rho}{\sqrt{6}\xi_1}(\partial_\mu \varphi)(\partial^\mu \rho)
-\frac{1}{2\xi_1}(\partial_\mu \rho)^2\, .
\end{align}
The potential of Eq. (\ref{vrphi}) is simplified as
\begin{align}
\label{h1-poten}
V_{h_1} = &\frac{\lambda_1+\lambda_2\rho^4 +2{\lambda}_L \rho^2}{8\xi_1^2} \,\left(1-e^{-2\varphi/\sqrt{6}}\right)^2  \, .
\end{align}
For a large non-minimal coupling, $\xi_1 \gg 1$, the kinetic mixing term $\partial_\mu \varphi \partial^\mu \rho$ can be neglected, and the kinetic
term for $\varphi$ field is approximately canonical. The kinetic term for $\rho$ field is non-canonical, and becomes canonically normalized 
by further field redefinition $\hat{\rho}=\frac{\rho}{\sqrt{\xi_1}}$ \cite{1105.2284}. In the large-field limit for inflation, we can obtain the mass of
the canonically normalized $\hat{\rho}$ from the potenial of Eq. (\ref{h1-poten}),
\beq
m^2_{\hat{\rho}} = \frac{\lambda_L}{2\xi_1}  \, ,
\eeq
which is requied to be larger than the Hubble parameter $H^2=\frac{\lambda_1}{\xi_1^2}$, leading to
\beq\label{mbigh1}
\lambda_L > \frac{2\lambda_1}{\xi_1}.
\eeq
The condition can be easily satisfied for a large $\xi_1$.

\item[(2)] $h_2$-inflation at $\rho^2=\infty$. Taking $\xi_1=0$ and replacing ($h_1$, $h_2$) with ($\rho$, $\varphi$), we can obtain the kinetic term from Eq. (\ref{eq-po0}) and Eq. (\ref{kine-h1h2}), 
\begin{align}
\label{h2-kin}
\calL_{kin-h_2} =&- \frac{1}{2}\left(1+\frac{\rho^2+1}{6\xi_2\rho^2}\right)(\partial_\mu \varphi)^2+\frac{1}{\sqrt{6}\xi_2\rho^3}(\partial_\mu \varphi)(\partial^\mu \rho)
-\frac{1}{2\xi_2\rho^4}(\partial_\mu \rho)^2\, .
\end{align}
The potential of Eq. (\ref{vrphi}) is simplified as
\begin{align}
\label{h2-poten}
V_{h_2} = &\frac{\lambda_1+\lambda_2\rho^4 +2{\lambda}_L \rho^2}{8\xi_2^2\rho^4} \,\left(1-e^{-2\varphi/\sqrt{6}}\right)^2  \, .
\end{align}
For $\xi_2 \gg 1$, the kinetic mixing term $\partial_\mu \varphi \partial^\mu \rho$ can be neglected, and the kinetic
term for $\varphi$ field is approximately canonical. The kinetic term for $\rho$ field becomes canonically normalized 
by further field redefinition $\hat{\rho}=\frac{1}{\sqrt{\xi_2}\rho}$ \cite{1105.2284}. The mass of
the canonically normalized $\hat{\rho}$ is obtained from the potenial of Eq. (\ref{h2-poten}),
\beq
m^2_{\hat{\rho}} = \frac{\lambda_L}{2\xi_2}  \, ,
\eeq
which is requied to be larger than the Hubble parameter $H^2=\frac{\lambda_2}{\xi_2^2}$, leading to
\beq\label{mbigh2}
\lambda_L > \frac{2\lambda_2}{\xi_2}.
\eeq
The condition can be easily satisfied for a large $\xi_2$.
\end{itemize}

Since the potential stability requires $\lambda_1>0$ and $\lambda_2>0$ as well as $\xi_1>0$ ($\xi_2>0$) for the $h_1$ ($h_2$)-inflation,
 Eq. (\ref{mbigh1}) and Eq. (\ref{mbigh2}) imply $\lambda_L>0$ for the $h_1$-inflation and the $h_2$-inflation.
The condition of $x_1>0$ and $x_2<0$ is naturally satisfied for the $h_1$-inflation with $\xi_2=0$. Similarly, 
$x_1<0$ and $x_2>0$ is satisfied for the $h_2$-inflation with $\xi_1=0$.

 The slow roll parameters used to characterize the inflation dynamics are,
 \beq\label{eq:sr}
\epsilon(\varphi) = \frac{1}{2} \left(\frac{dV(\varphi)/d\varphi}{V(\varphi)} \right)^2 \, , 
\qquad \eta(\varphi) =  \frac{d^2V(\varphi)/d\varphi^2}{V(\varphi)} \, .
\eeq
The field value at the end of inflation $\varphi_{\rm e}$ is determined by $\epsilon= 1$, and the horizon exit value $\varphi_{\rm *}$ can be calculated by assuming
an e-folding number between the two periods,
\beq
N = \int_{\varphi_{\rm e}}^{\varphi_{\rm *}} d\varphi \frac{1}{\sqrt{2 \epsilon}}  \, .
\eeq
Taking $N=$ 60 and the slow roll parameters at the $\varphi_*$, we calculate the inflationary observables of 
spectrum index $n_s$, the tensor to scalar ratio of $r$, and the scalar amplitude $P_s$,
\begin{align}\label{eq:delta}
n_s  &= 1 + 2 \,\eta - 6\, \epsilon ~|_{\varphi_*}=0.9678\, ,\\
r &= 16\, \epsilon ~|_{\varphi_*}=0.003\, ,\\
P_s & =\frac{V}{24\pi^2\epsilon} ~|_{\varphi_*}.
\end{align}

The Planck collaboration reported constraints on the three inflation observables \cite{pl2018}: 
\begin{align}
 n_s &= 0.9649\pm 0.0042,\\
 r &< 0.056,\\
 P_s &= (2.099\pm 0.014)\times 10^{-9}.\label{tensca}
\end{align}
The values of $n_s$ and $r$ are well consistent with the Plank bounds. The scalar amplitude $P_s$ imposes
very stringent constraint on $\lambda_1$ and $\xi_1$ for $h_1$-inflation and $\lambda_2$ and $\xi_2$ for $h_2$-inflation.

\section{Inflation and relevant constraints at low energy}
\subsection{Numerical calculations}
We take the light CP-even Higgs boson $h$ as the
SM-like Higgs, $m_h=125$ GeV. 
The experimental data of $b \to s\gamma$ imposes stringent bound on the charged Higgs mass of the type-II 2HDM, 
$m_{H^{\pm}} > 570$ GeV \cite{bsr570-exp,bsr570-th}.
In the type-I model the bound of $b \to s\gamma$ can be sizably alleviated, and we take $m_{H^{\pm}} > 80$ GeV considering
the constraints from search for Higgs at the LEP collider. The other key parameters are scanned over in the following ranges:
\begin{align}
&150 {\rm GeV} < m_H < 2000 {\rm GeV},~~ m_A > 10 {\rm GeV},~~0.93<\mid\sin(\beta-\alpha)\mid < 1,\nonumber\\
&1< \tan\beta < 50 {\rm~for~type-I},~~1< \tan\beta < 15 {\rm~for~type-II}.
\label{epot2}
\end{align}

To perform the inflation condition, we run $\lambda_i$ and $m_{ij}$ from the electroweak scale to unitarity scale $\mu_U = min(1/\xi_1, 1/\xi_2)$ via the two-loop renormalization group (RG) equations, which is implemented by $\textsf{2HDME}$ \cite{2hdme}. 
The theory is only well defined up to the scale $\mu_U = min(1/\xi_1, 1/\xi_2)$ at which unitarity is violated by
 the $WW$ scattering processes with exchange of graviton \cite{uv-1,uv-2,1310.7410,0903.0355,0912.5463,1002.2995,1008.5157,1403.3219,1404.4627}. 
Therefore, additional new physics should be introduced at the unitarity scale $\mu_U$ to restore unitarity, which is beyond
the scope of this paper. The new physics is generally assumed to not significantly affect the discussions in the previous section, 
but it is likely to affect the running of the relevant parameters above $\mu_U$ \cite{1105.2284}.
Due to uncertainties from new physics above $\mu_U$, the running of the parameters during inflation
 cannot be reliably calculated \cite{run-uv-1,run-uv-2,run-uv-3,run-uv-4}.
 Therefore, the running of the parameters above the unitarity scale $\mu_U$ are omitted, see e.g. \cite{infla-inert1,infla-g2hdm}.

We impose the following theoretical constraints:
\begin{itemize}
\item[(1)] Perturbativity. To satisfy the perturbativity during the RG evolution, we impose the 
upper limit on the quartic coupling,
\beq
\lambda_i(Q)\leq 4\pi.
\eeq

\item[(2)] Vacuum stability. In order to ensure vacuum stability, the potential should be positive for large values of
the fields, which requires
\bea
&&\lambda_1(Q) > 0,~\lambda_2(Q) > 0,~ \lambda_3(Q)+\sqrt{\lambda_1(Q)\lambda_2(Q)} > 0,\nonumber\\
&&\lambda_3(Q)+\lambda_4(Q)-\mid\lambda_5(Q)\mid+\sqrt{\lambda_1(Q)\lambda_2(Q)} > 0.
\eea
Here we take the RG improved potential where the parameters are
replaced by their two-loop running couplings. Taking the type-II 2HDM as an example, 
the full one-loop effective potential can revive a large fraction of points which is ruled out by the vacuum stability of the pure tree-level potential \cite{sta-1loop-1,sta-1loop-2}. However, the results of the vacuum stability of the one-loop effective potential are essentially 
in agreement with the RG improved potential with the two-loop running couplings \cite{sta-1loop-1}.  

\item[(3)] Unitarity. To respect unitarity, the eigenvalues of the $2\to 2$ scalar scattering matrix are imposed the following
bounds \cite{unit-2h1,unit-2h2},
\begin{eqnarray}
 && \tfrac{3}{2}(\lambda_1(Q)+\lambda_2(Q)) \pm \sqrt{\tfrac{9}{4}(\lambda_1(Q)-\lambda_2(Q))\raisebox{0.3pt}{$^2$}+(2\lambda_3(Q)+\lambda_4(Q))^2}\leq 8\pi \,, \nonumber\\
&& \tfrac{1}{2}(\lambda_1(Q)+\lambda_2(Q)) \pm
\sqrt{\tfrac{1}{4}(\lambda_1(Q)-\lambda_2(Q))\raisebox{0.3pt}{$^2$}+\lambda_4(Q)^2}\leq 8\pi \,, \nonumber\\
&& \tfrac{1}{2}(\lambda_1(Q)+\lambda_2(Q)) \pm
\sqrt{\tfrac{1}{4}(\lambda_1(Q)-\lambda_2(Q))\raisebox{0.3pt}{$^2$}+\lambda_5(Q)^2}\leq 8\pi \,, \nonumber\\
&& \lambda_3(Q) + 2 \lambda_4(Q) \pm 3 \lambda_5(Q)\leq 8\pi \,, \nonumber\\
&&\, \lambda_3(Q) \pm \lambda_4(Q) \leq 8\pi\,,\nonumber\\
&&\, \lambda_3(Q) \pm \lambda_5(Q) \leq 8\pi\,.
\end{eqnarray}
\end{itemize}

In the following discussions, $\lambda_i$ is used to denote the quartic coupling at electroweak scale, and the corresponding quartic coupling at 
unitarity scale is expressed by $\bar{\lambda}_i$.
In addition to the inflation condition and theoretical constraints, we consider the following observables at the low energy:

\begin{itemize}
\item[(1)] The oblique parameters. The $S$, $T$, $U$ parameters can give strong constraints on 
the Higgs mass spectrum of 2HDM. We employ $\textsf{2HDMC}$ \cite{2hc-1} to calculate the $S$, $T$, $U$ parameters, and
fit the results of Ref. \cite{pdg2018},
\beq
S=0.02\pm 0.10,~~  T=0.07\pm 0.12,~~ U=0.00 \pm 0.09,
\eeq
with the correlation coefficients of
\beq
\rho_{ST} = 0.92, ~~\rho_{SU} = -0.66, ~~\rho_{TU} = -0.86.
\eeq

\item[(2)] The flavor observables and $R_b$. $\textsf{SuperIso-3.4}$ \cite{spriso} is used to 
calculate $Br(B\to X_s\gamma)$, and $\Delta m_{B_s}$ is calculated following the
formulas in \cite{deltmq}. Besides, following the formulas in \cite{rb1,rb2} we calculate $R_b$ of bottom quarks produced in $Z$ decays.

\begin{table}
\begin{footnotesize}
\begin{tabular}{| c | c | c | c |}
\hline
\textbf{Channel} & \textbf{Experiment} & \textbf{Mass range [GeV]}  &  \textbf{Luminosity} \\
\hline
{$gg/b\bar{b}\to H/A \to \tau^{+}\tau^{-}$} & CMS 13 TeV \cite{1709.07242}& 200-2250   & 36.1 fb$^{-1}$ \\
{$gg/b\bar{b}\to H/A \to \tau^{+}\tau^{-}$} & ATLAS 13 TeV \cite{2002.12223}& 200-2500   & 139 fb$^{-1}$ \\
\hline
{$gg\to H/A \to t\bar{t}$} & CMS 13 TeV \cite{1908.01115}& 400-750   & 35.9 fb$^{-1}$ \\
\hline
{$gg\to H/A \to \gamma\gamma$~+~$t\bar{t}H/A~(H/A\to \gamma\gamma)$}& CMS 13 TeV \cite{HIG-17-013-pas}& 70-110 & 35.9 fb$^{-1}$ \\
{$VV\to H \to \gamma\gamma$~+~$VH~(H\to \gamma\gamma)$}& CMS 13 TeV \cite{HIG-17-013-pas}& 70-110 & 35.9 fb$^{-1}$ \\
\hline

{$gg/VV\to H\to W^{+}W^{-}~(\ell\nu qq)$} & ATLAS 13 TeV  \cite{1710.07235}& 200-3000  &  36.1 fb$^{-1}$\\
{$gg/VV\to H\to W^{+}W^{-}~(e\nu \mu\nu)$} & ATLAS 13 TeV  \cite{1710.01123}& 200-3000  &  36.1 fb$^{-1}$\\
{$gg/VV\to H\to W^{+}W^{-}$} & CMS 13 TeV  \cite{1912.01594}& 200-3000  &  35.9 fb$^{-1}$\\
\hline
$gg/VV\to H\to ZZ $ & ATLAS 13 TeV~\cite{1712.06386} & 200-2000 & 36.1 fb$^{-1}$ \\
$gg/VV\to H\to ZZ $ & ATLAS 13 TeV~\cite{1708.09638} & 300-5000 & 36.1 fb$^{-1}$ \\
$gg/VV\to H\to ZZ $ & ATLAS 13 TeV~\cite{2009.14791} & 200-2000 & 139 fb$^{-1}$ \\
\hline

\end{tabular}
\end{footnotesize}
\caption{The upper limits at 95\%  C.L. on the production cross-section times branching ratio of
$\tau^+\tau^-$, $t\bar{t}$, $\gamma\gamma$, $WW$, and $ZZ$ considered in 
the $H$ and $ A $ searches at the LHC.}
\label{tabh}
\end{table}

\begin{table}
\begin{footnotesize}
\begin{tabular}{| c | c | c | c |}
\hline
\textbf{Channel} & \textbf{Experiment} & \textbf{Mass range [GeV]}  &  \textbf{Luminosity} \\
\hline
$gg \to H\to hh \to b\bar{b}b\bar{b}$ & CMS 13 TeV~\cite{1710.04960} & 750-3000  &  35.9 fb$^{-1}$ \\
$gg \to H\to hh \to (b\bar{b}) (\tau^{+}\tau^{-})$ & CMS 13 TeV~\cite{1707.02909} & 250-900  &  35.9 fb$^{-1}$ \\
$pp \to H\to hh $ & CMS 13 TeV~\cite{1811.09689} & 250-3000  &  35.9 fb$^{-1}$ \\
$gg \to H\to hh \to b\bar{b}ZZ$ & CMS 13 TeV~\cite{2006.06391} & 260-1000  &  35.9 fb$^{-1}$ \\

$gg \to H\to hh \to b\bar{b}\tau^{+}\tau^{-}$ & CMS 13 TeV~\cite{2007.14811} & 1000-3000  &  139 fb$^{-1}$ \\
\hline


{$gg/b\bar{b}\to A\to hZ\to (b\bar{b})Z$}& ATLAS 13 TeV \cite{1712.06518}& 200-2000 & 36.1 fb$^{-1}$  \\

{$gg/b\bar{b}\to A\to hZ\to (b\bar{b})Z$}& CMS 13 TeV \cite{1903.00941}& 225-1000 & 35.9 fb$^{-1}$  \\

{$gg\to A\to hZ\to (\tau^{+}\tau^{-}) (\ell \ell)$}& CMS 13 TeV \cite{1910.11634}& 220-400 & 35.9 fb$^{-1}$  \\
\hline


{$pp\to  h \to AA \to (b\bar{b})(\tau^{+}\tau^{-})$} & CMS 13 TeV~\cite{1805.10191} &  15-60  &35.9 fb$^{-1}$ \\
{$pp\to  h \to AA \to \tau^{+}\tau^{-}\tau^{+}\tau^{-}$} & CMS 13 TeV~\cite{1907.07235} &  4-15  &35.9 fb$^{-1}$ \\
{$pp\to  h \to AA \to \mu^{+}\mu^{-}\tau^{+}\tau^{-}$} & CMS 13 TeV~\cite{2005.08694} &  3.6-21  &35.9 fb$^{-1}$ \\
\hline
$gg/b\bar{b}\to A(H)\to H(A)Z\to (b\bar{b}) (\ell \ell)$ & ATLAS 13 TeV \cite{1804.01126}& 130-800 & 36.1 fb$^{-1}$ \\

$gg\to A(H)\to H(A)Z\to (b\bar{b}) (\ell \ell)$ & CMS 13 TeV \cite{1911.03781}& 30-1000 & 35.9 fb$^{-1}$ \\
\hline
\end{tabular}
\end{footnotesize}
\caption{The upper limits at 95\%  C.L. on the production cross-section times branching ratio for the channels
of Higgs-pair and a Higgs production in association with $Z$ at the LHC.}
\label{tabhh}
\end{table}

\item[(3)] The global fit to the 125 GeV Higgs signal data. We employ the version 2.0 of $\textsf{Lilith}$ \cite{lilith-1,lilith-2} to perform the
$\chi^2$ calculation for the signal strengths of the 125 GeV Higgs combining the LHC run-I and
run-II data. We require $\chi^2-\chi^2_{\rm min} \leq 6.18$ with $\chi^2_{\rm min}$
being the minimum of $\chi^2$. These surviving samples mean to be within
the $2\sigma$ range in any two-dimension plane of the
model parameters explaining the Higgs data.

\item[(4)] The exclusion limits of searches for additional Higgs bosons. $\textsf{HiggsBounds-4.3.1}$ \cite{hb1,hb2} is employed to perform the exclusion
constraints from the neutral and charged Higgs searches at LEP at 95\% confidence level.

We use $\textsf{SusHi}$ to calculate the cross sections for $H$ and $A$ in the
gluon fusion and $b\bar{b}$-associated production at NNLO in QCD \cite{sushi}. 
The cross sections of $H$ via vector boson fusion process are deduced from results of the LHC
Higgs Cross Section Working Group \cite{higgswg}.
The top quark loop and $b$-quark loop respectively have destructive and constructive interference contributions to $gg\to A$ production 
in the type-II and type-I 2HDMs. Therefore, the contributions of top quark loop always dominate over those of $b$-quark loop in
 the type-I model. In the type-II model, the cross section of $gg\to A$ decreases with an increase
of $\tan\beta$, reaches the minimum value for the moderate $\tan\beta$, and is dominated by the $b$-quark loop for enough large
 $\tan\beta$ \cite{1312.4759}. The cross section of $gg\to H$ depends on $\sin(\beta-\alpha)$ in addition to $\tan\beta$
and $m_H$. $\textsf{2HDMC}$ is employed to calculate the branching ratios of the
various decay modes of $H$ and $A$.

We consider the searches for additional Higgs bosons at LHC, including $h\to AA$, $H/A\to \tau^+\tau^-,t\bar{t}$, $H/A\to \gamma\gamma,~VV$,
$H\to hh$, $A\to hZ,~HZ$, $H\to AZ$. In Tables \ref{tabh} and \ref{tabhh}, we list the ATLAS and CMS analyses at the 13 TeV LHC with more than 
35.9 fb$^{-1}$ integrated luminosity data. The analyses at the 8 TeV LHC and 13 TeV with less than 35.9 fb$^{-1}$ integrated luminosity data are 
also included, which may be found in Ref. \cite{pt2h-lwang}.
\end{itemize}

\subsection{Results and discussions}
\subsubsection{Higgs inflation in type-I 2HDM}
\begin{figure}[tb]
\centering
 \epsfig{file=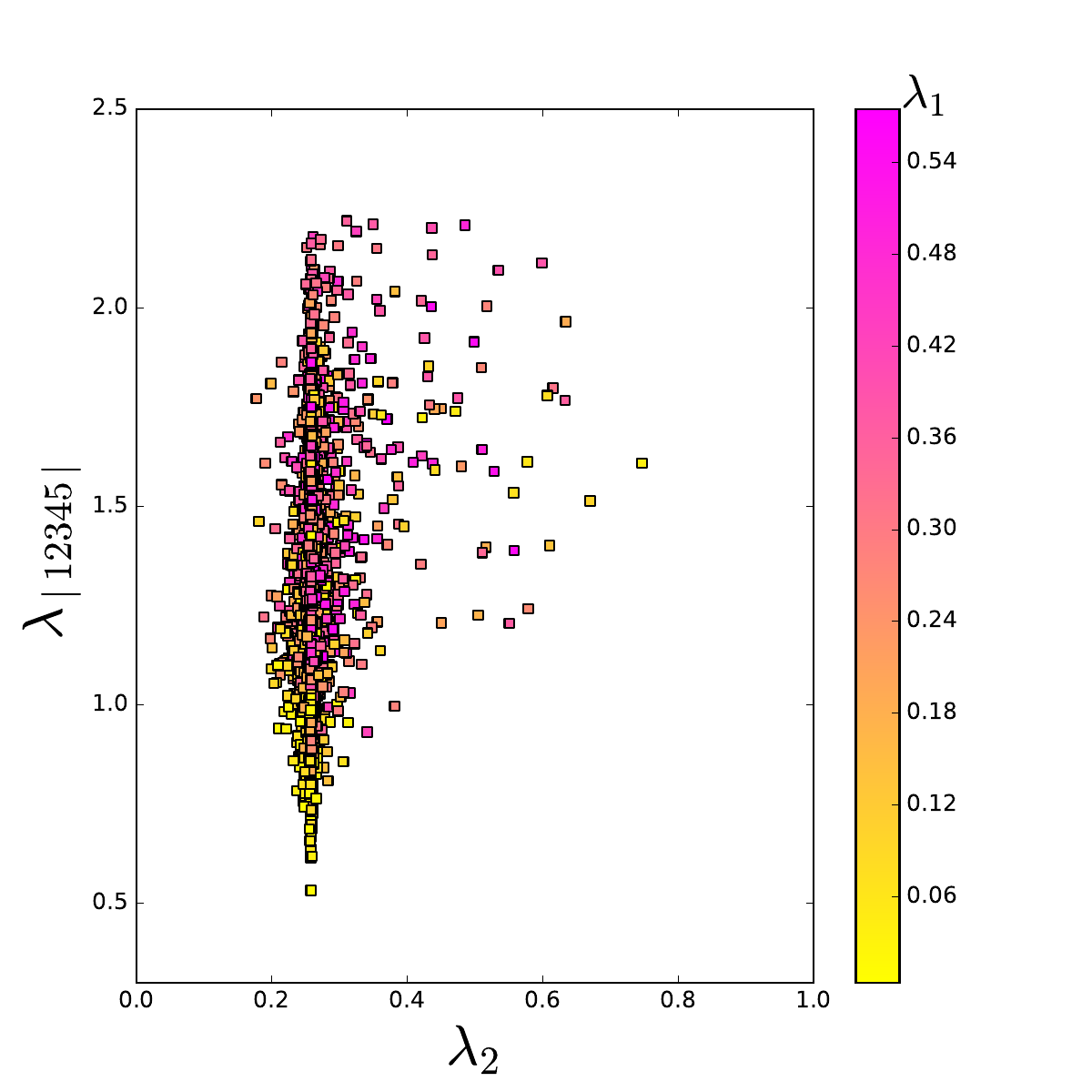,height=8.0cm}
\vspace{-0.0cm} \caption{For the type-I model, scatter plots of $\lambda_{\mid 12345 \mid}$ and $\lambda_2$ satisfying the constraints of pre-inflation and $h_2$-inflation.} \label{t1rilam2}
\end{figure}

\begin{figure}[tb]
\centering
 \epsfig{file=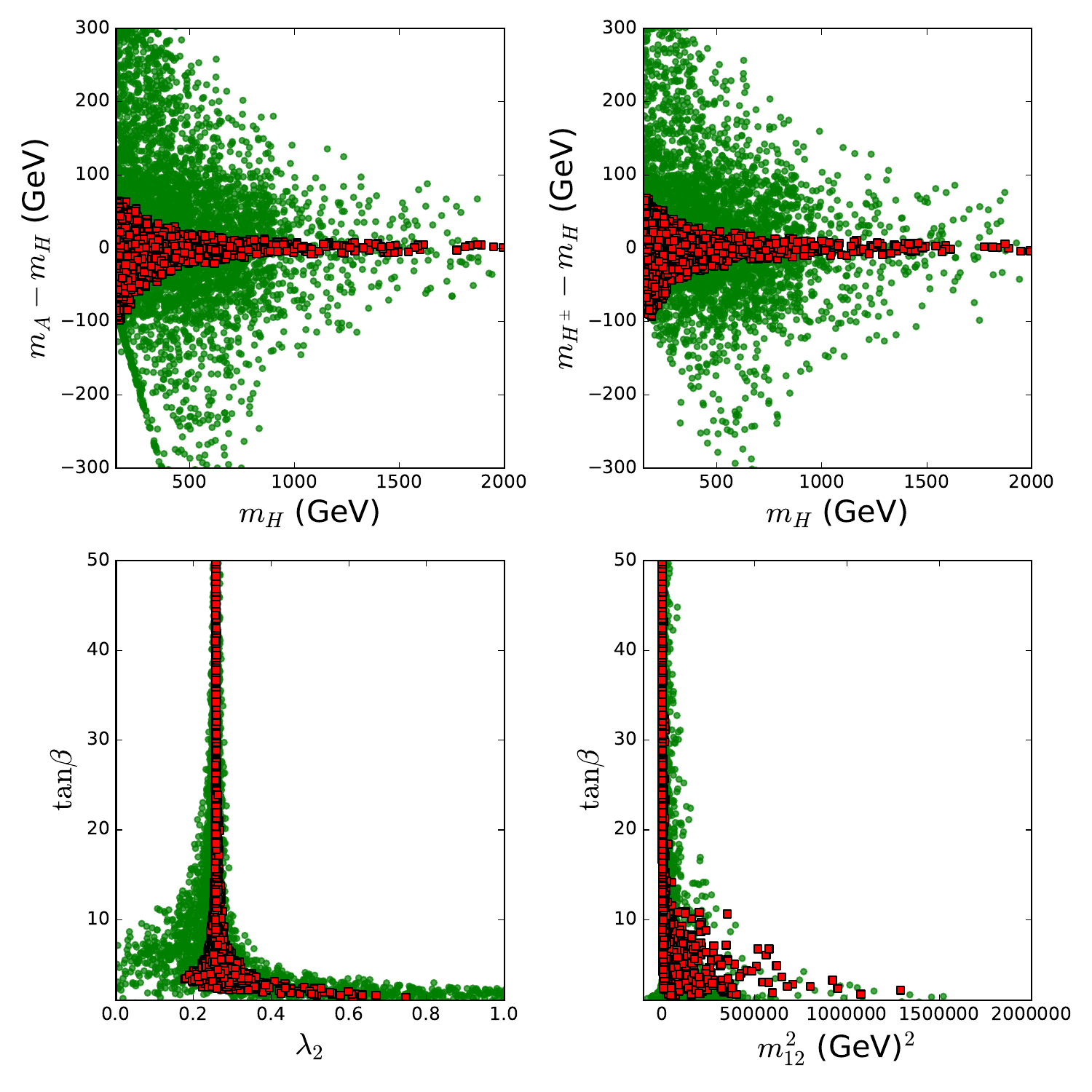,height=15cm}
\vspace{-0.2cm} \caption{For the type-I model, the bullets (green) satisfying the oblique parameters, the signal data of 
the 125 GeV Higgs, and theoretical constraints at the electroweak scale. The squares (red) satisfying the constraints of 
pre-inflation and $h_2$-inflation.} \label{t1ridm}
\end{figure}

In Fig. \ref{t1rilam2}, we impose the constraints of "pre-inflation" (denoting theoretical constraints, the oblique parameters, the signal data of 
the 125 GeV Higgs), and show the surviving samples achieving the $h_2$-inflation in the type-I 2HDM. The vacuum stability,
perturbativity, and unitarity impose a stringent upper bound on $\lambda_{\mid 12345 \mid}$ 
( $\lambda_{\mid 12345\mid}\equiv\mid\lambda_1\mid+\mid\lambda_2\mid+\mid\lambda_3\mid+\mid\lambda_4\mid+\mid\lambda_5\mid$) at the electroweak scale, $\lambda_{\mid 12345 \mid}<$ 2.5. 
If the quartic couplings at electroweak scale is too large, although they may satisfy the theoretical constraints at electroweak scale, 
the theoretical constraints at the high energy scale will be not satisfied with their evolution.  

In Fig. \ref{t1ridm}, we comparatively show the surviving samples satisfying the oblique parameters, the signal data of 
the 125 GeV Higgs, and theoretical constraints at the electroweak scale, and those satisfying 
the constraints of pre-inflation and $h_2$-inflation.
From Fig. \ref{t1ridm}, we find that the inflation can be achieved in whole range of 150 GeV $<m_H<$ 2 TeV, and
 favors a small mass splitting between $m_A~(m_{H^\pm})$ and $m_H$. The mass splitting is favored to decrease with an increase of $m_H$.
$m_A~(m_{H^\pm})-m_H$ is favored to vary from -100 GeV to 70 GeV for a small $m_H$, and $H,~A,~H^\pm$ tend to be nearly degenerate in mass for $m_H$ approaching
to 2 TeV. Schematically, the squared masses of $H,~ A$, and $H^\pm$ can be given as \cite{0408364,1609.04185},
\beq\label{mlami}
m_\phi^2\approx y_\phi M^2  + f_\phi(\lambda_i) v^2 + \ord(\frac{v^4}{M^2}),
\eeq
where $M^2=\frac{m_{12}^2}{c_\beta s_\beta}$ with $s_\beta\equiv\sin\beta$ and 
$c_\beta\equiv\cos\beta$. $y_\phi=1$ for $A,~H^\pm$, and $y_\phi=\sin(\beta-\alpha)$ for $H$. Since
 the theoretical constraints require small $\lambda_i$, which lead to small mass splitting between $H,~ A,$ and $H^\pm$ according to
Eq. (\ref{mlami}).

The lower panel of  Fig. \ref{t1ridm} show that $m_{12}^2$ is favored to have small value and $\lambda_2$ is favored to be around 0.26 
for a large $\tan\beta$. The main reason is from the theoretical constraints.
The vacuum stability requires,
\beq
\lambda_1>0,~~\lambda_2>0,~~\lambda_3>-\sqrt{\lambda_1\lambda_2}\,,~~\lambda_3+\lambda_4-\mid\lambda_5\mid>-\sqrt{\lambda_1\lambda_2}\,,
\label{vaceq}
\eeq
with \cite{0207010,ws-9-2}
\begin{eqnarray}\label{poten-cba}
 &&v^2 \lambda_1  = \frac{m_H^2 c_\alpha^2 + m_h^2 s_\alpha^2 - m_{12}^2 t_\beta}{ c_\beta^2}, \ \ \ 
v^2 \lambda_2 = \frac{m_H^2 s_\alpha^2 + m_h^2 c_\alpha^2 - m_{12}^2 t_\beta^{-1}}{s_\beta^2},  \nonumber \\  
&&v^2 \lambda_3 =  \frac{(m_H^2-m_h^2) s_\alpha c_\alpha + 2 m_{H^{\pm}}^2 s_\beta c_\beta - m_{12}^2}{ s_\beta c_\beta }, \ \ \ 
v^2 \lambda_4 = \frac{(m_A^2-2m_{H^{\pm}}^2) s_\beta c_\beta + m_{12}^2}{ s_\beta c_\beta },  \nonumber \\
 &&v^2 \lambda_5=  \frac{ - m_A^2 s_\beta c_\beta  + m_{12}^2}{ s_\beta c_\beta }\, , 
 \label{eq:lambdas}
\end{eqnarray}
where $t_\beta\equiv\tan\beta$, $c_\alpha\equiv\cos\alpha$, and $s_\alpha\equiv\sin\alpha$.
 When $\sin(\beta-\alpha)$ is very closed to 1.0, we can approximately obtain the following relations,
\begin{eqnarray}
v^2 \lambda_1 &=&  m_h^2 - \frac{t_\beta^3\,(m_{12}^2 -m_H^2  s_\beta c_\beta ) }{ s_\beta^2}\,,\nonumber \\
v^2 \lambda_2 &=& m_h^2 - \frac{ (m_{12}^2 -m_H^2  s_\beta c_\beta) }{ t_\beta s_\beta^2 }\,,\nonumber\\
\label{eq:alignment2_lambda}
v^2 \lambda_3 &=&  m_h^2 + 2 m_{H^{\pm}}^2 - 2m_H^2 -  \frac{t_\beta (m_{12}^2 -m_H^2  s_\beta c_\beta)}{  s_\beta^2}\,,\nonumber\\
v^2 \lambda_4 &=&  m_A^2-  2 m_{H^{\pm}}^2 + m_H^2+  \frac{t_\beta (m_{12}^2 -m_H^2  s_\beta c_\beta)}{  s_\beta^2}\,,\nonumber\\
v^2 \lambda_5 &=&  m_H^2 - m_A^2+  \frac{t_\beta (m_{12}^2 -m_H^2  s_\beta c_\beta)}{  s_\beta^2} \,.
\label{poten-cba0}
\end{eqnarray}
For a large $\tan\beta$, the first condition of Eq. (\ref{vaceq}) favors $m^2_{12}- m^2_H s_\beta c_\beta$ $\to$ 0.
As a result, we deduce $\lambda_2\approx0.26$ from the second equation of Eq. (\ref{poten-cba0}).

\begin{figure}[tb]
\centering
 \epsfig{file=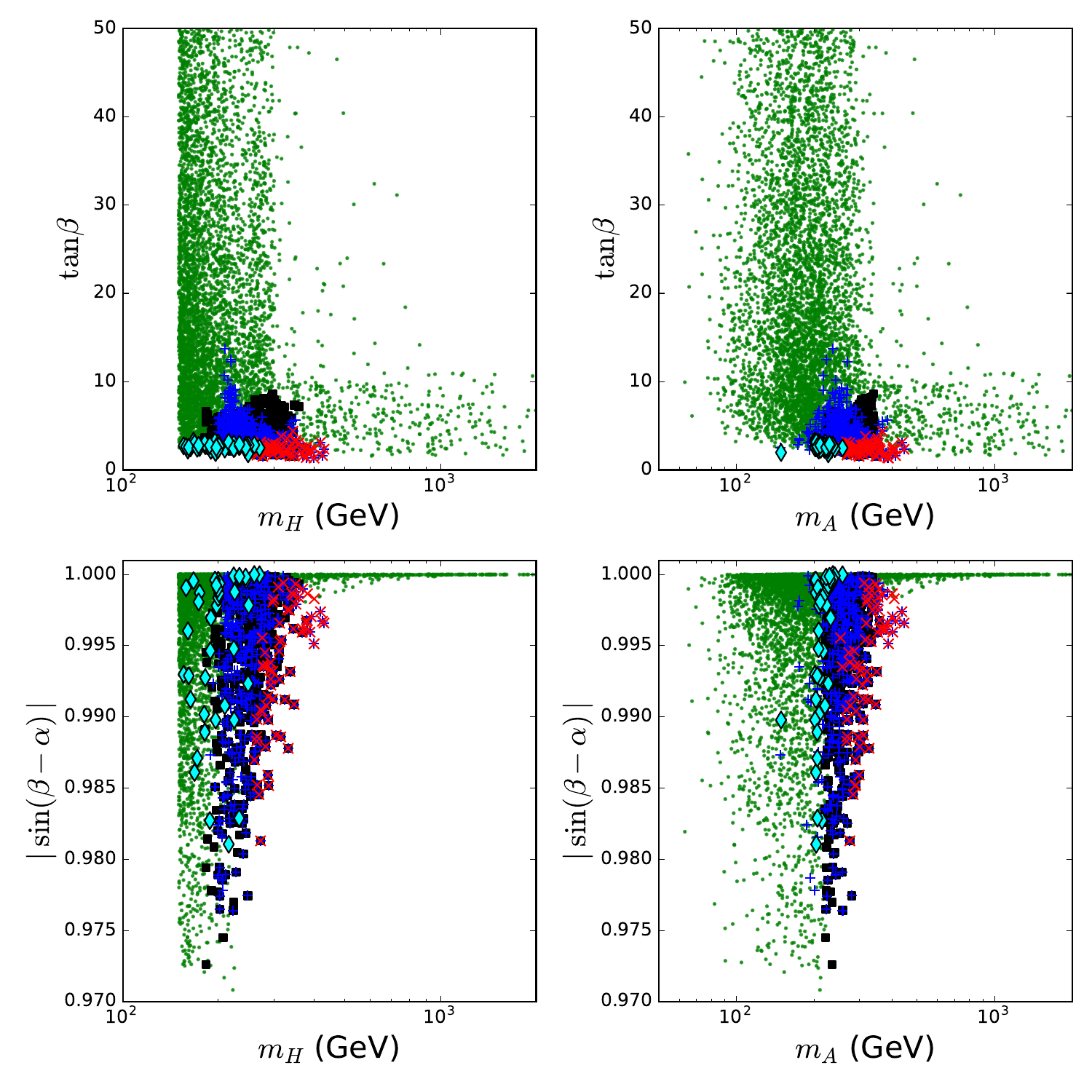,height=15cm}
\vspace{-0.5cm} \caption{In the type-I model, all the surviving samples satisfies the constraints of pre-inflation, $h_2$-inflation, 
the exclusion limits from searches for Higgs at LEP, the flavor observables and $R_b$. 
The pluses (blue), diamonds (cyan), crosses (red), and squares (black) are respectively excluded by $H/A\to VV, \gamma\gamma$, $H/A\to \tau^+\tau^-$, 
$H\to hh$, and $A\to hZ$ channels at the LHC. The bullets (green) are
allowed by various LHC direct searches.} \label{t1rilhc}
\end{figure}
\begin{figure}
\centering
 \epsfig{file=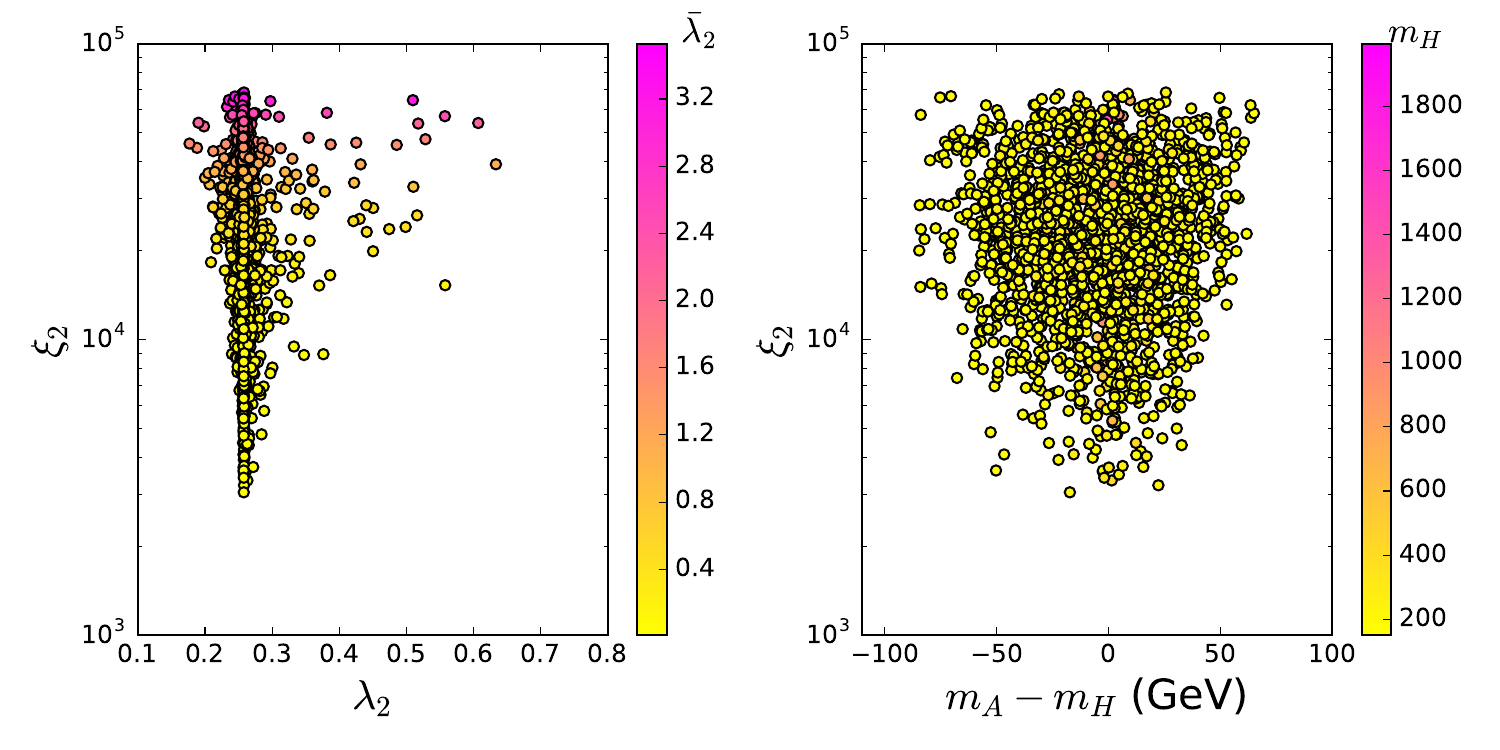,height=7cm}
\vspace{-0.6cm} \caption{In the type-I model, scatter plots of $\xi_2$ versus $\lambda_2$ and $\xi_2$ versus $m_A-m_H$ satisfying the constraints of
pre-inflation, $h_2$-inflation, the flavor observables, the exclusion limits from searches for Higgs at LEP, the flavor observables, $R_b$, and the direct searches at the LHC.} \label{t1rixi2}
\end{figure}

\begin{figure}
\centering
 \epsfig{file=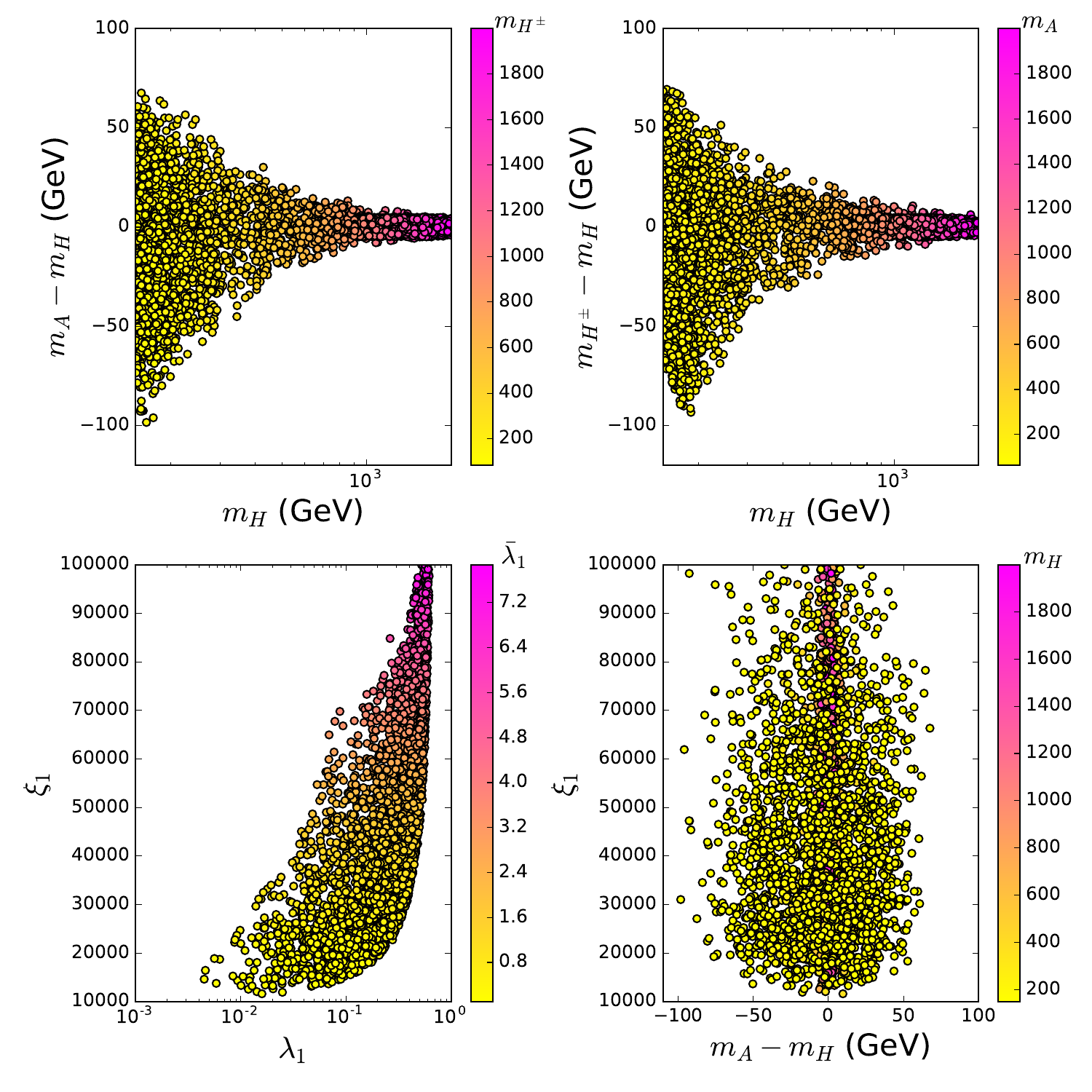,height=13cm}
\vspace{-0.6cm} \caption{In the type-I model, all the surviving samples satisfies the constraints of pre-inflation, 
$h_1$-inflation, the exclusion limits from searches for Higgs at LEP, the flavor observables, $R_b$, and the direct searches at the LHC.} \label{t1r0xi1}
\end{figure}

\begin{figure}[tb]
\centering
 \epsfig{file=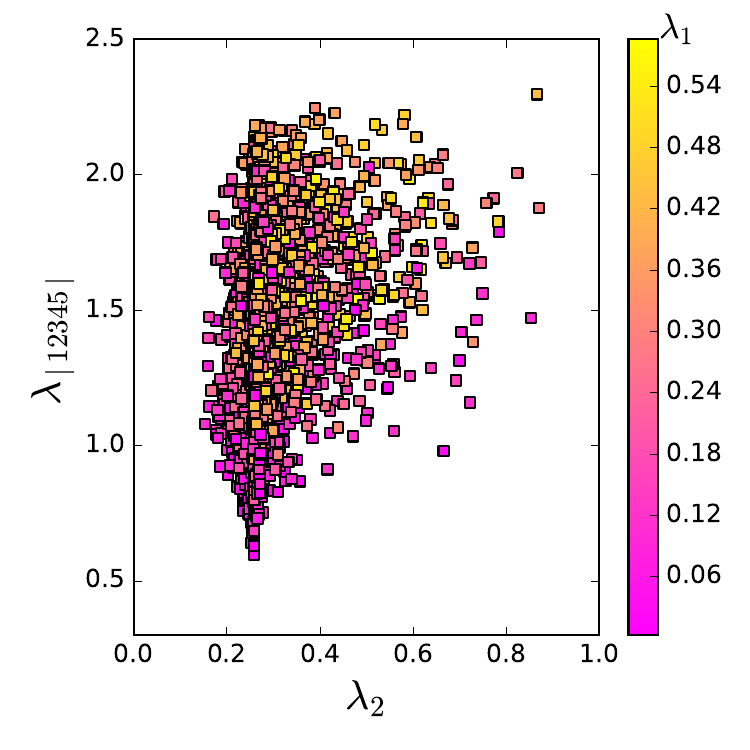,height=7cm}
\vspace{-0.6cm} \caption{For the type-II model, scatter plots of $\lambda_{\mid 12345 \mid}$ and $\lambda_2$ satisfying the constraints of pre-inflation 
and $h_2$-inflation.} \label{t2rilam2}
\end{figure}

\begin{figure}[tb]
\centering
 \epsfig{file=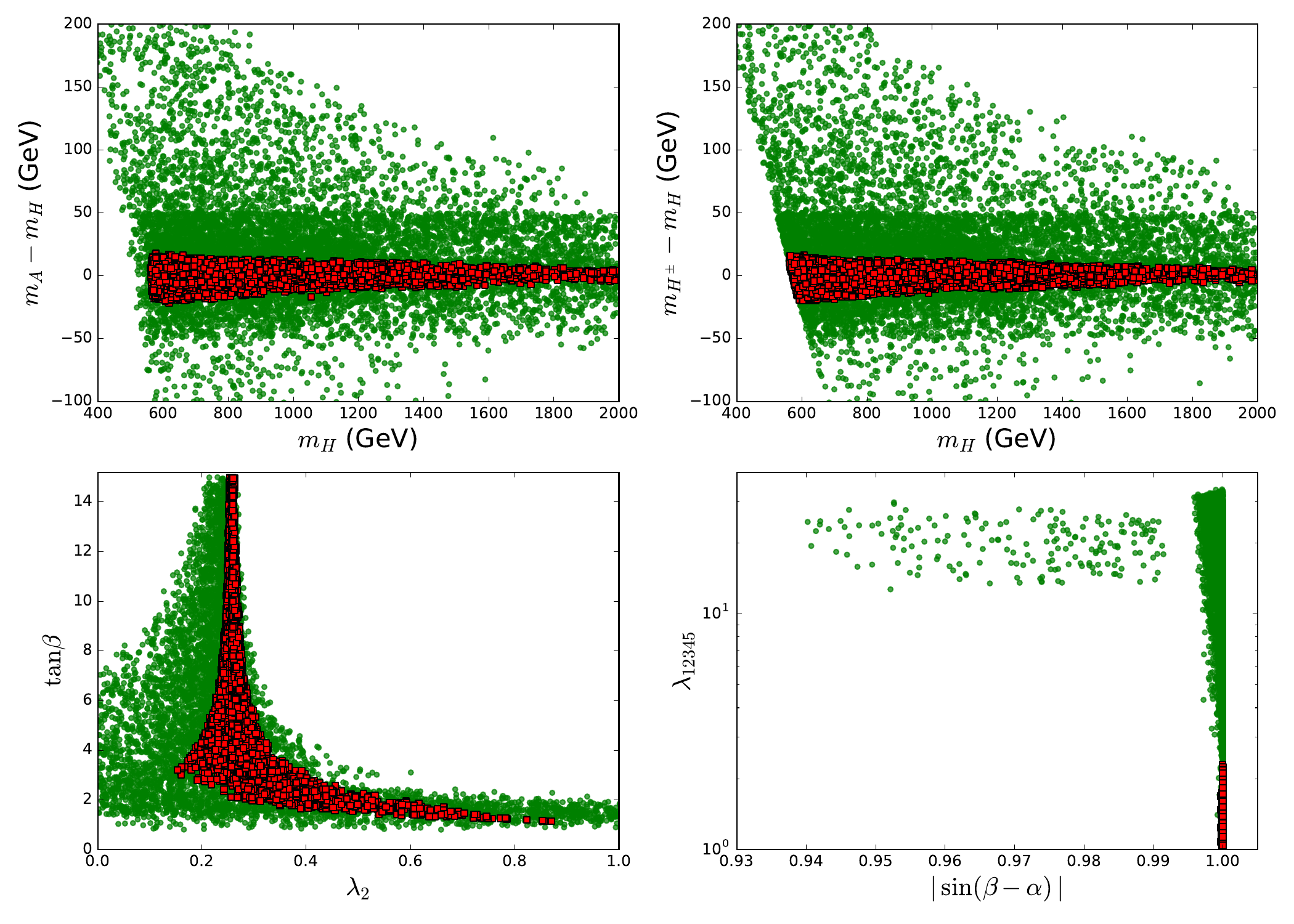,height=10cm}
\vspace{-0.6cm} \caption{For the type-II model, the bullets (green) satisfying the oblique parameters, the signal data of 
the 125 GeV Higgs, and theoretical constraints at the electroweak scale. The squares (red) satisfying the constraints of 
pre-inflation and $h_2$-inflation.} \label{t2ridm}
\end{figure}

\begin{figure}[tb]
\centering
 \epsfig{file=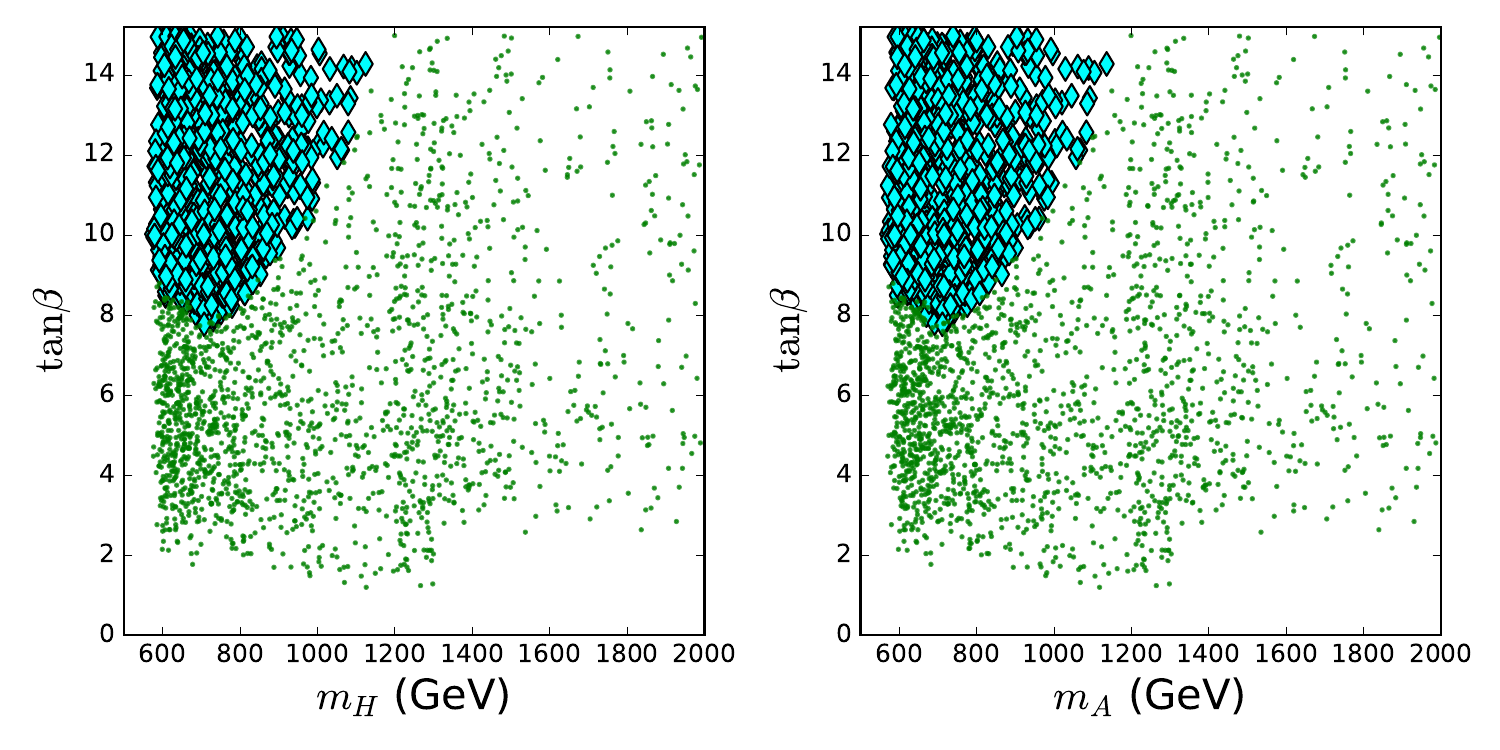,height=6cm}
\vspace{-0.5cm} \caption{In the type-II model, all the surviving samples satisfies the constraints of pre-inflation, $h_2$-inflation, 
the exclusion limits from searches for Higgs at LEP, the flavor observables, and $R_b$. 
The diamonds (cyan) are excluded by $H/A\to \tau^+\tau^-$ channels at the LHC, and the bullets (green) are
allowed by various LHC direct searches.} \label{t2rilhc}
\end{figure}

\begin{figure}[tb]
\centering
 \epsfig{file=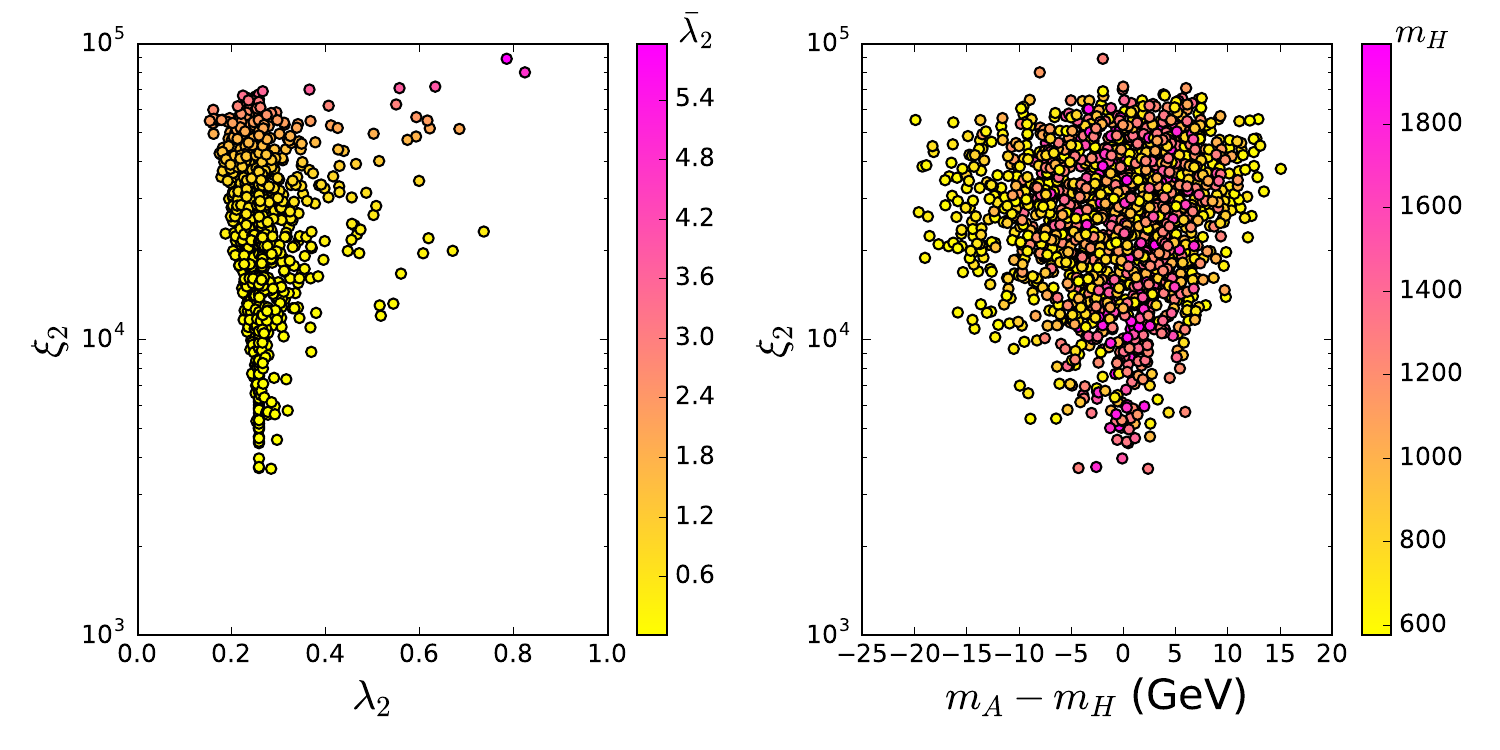,height=7cm}
\vspace{-0.5cm} \caption{In the type-II model, scatter plots of $\xi_2$ and $\lambda_2$ satisfying the constraints of
pre-inflation, $h_2$-inflation, the exclusion limits from searches for Higgs at LEP, the flavor observables, $R_b$, and the direct searches at the LHC.} \label{t2rixi2}
\end{figure}

\begin{figure}[tb]
\centering
 \epsfig{file=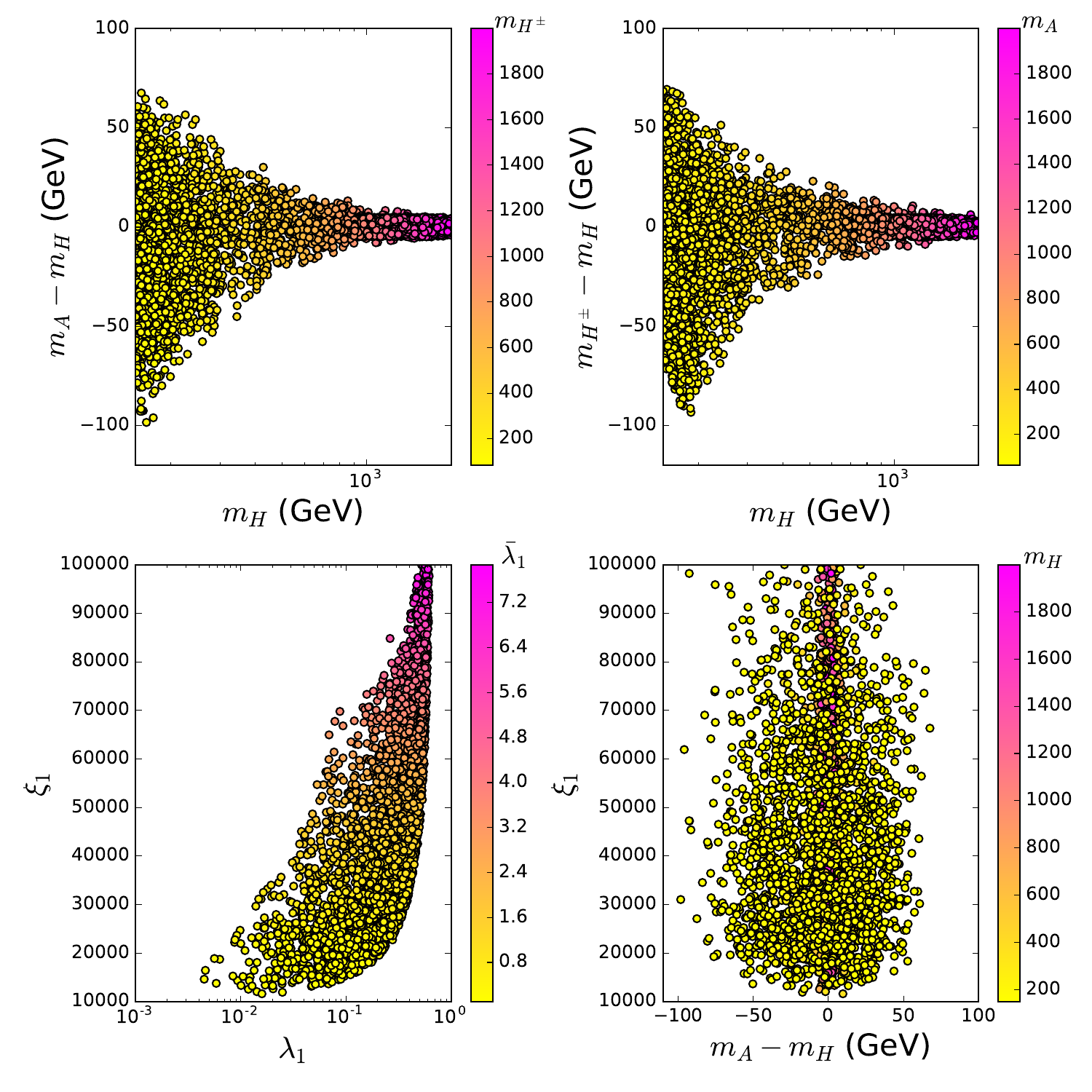,height=13cm}
\vspace{-0.5cm} \caption{In the type-II model, all the surviving samples satisfies the constraints of pre-inflation, 
$h_1$-inflation, the exclusion limits from searches for Higgs at LEP, the flavor observables, $R_b$, and the direct searches at the LHC.} \label{t2r0xi1}
\end{figure}

In Fig. \ref{t1rilhc}, we show the constraints of the direct searches for Higgs bosons at the LHC on the samples achieving the $h_2$-inflation
 in the type-I model. The $H/A\to VV, \gamma\gamma$, $H/A\to \tau^+\tau^-$, $H\to hh$, and $A\to hZ$ channels can exclude many 
points achieving $h_2$-inflation in
the region of $m_A$ ($m_H$) $<450$ GeV. However, $\tan\beta$ is allowed to take large enough to suppress the production cross sections of
$H$ and $A$ at the LHC. As a result, the constraints of these channels can be satisfied for a large $\tan\beta$. Since the signal data of the 125 GeV Higgs
impose very stringent bound on the $hAA$ coupling, the direct searches for $h\to AA$ channels at the LHC fail to constrain the parameter space.
Because we take $\tan\beta > $ 1, $H/A\to t\bar{t}$ do not impose any constraints. The inflation favors a small mass splitting between $H$ and $A$, which
leads that there are no constraints from $H\to AZ$ and $A\to HZ$ channels.

After imposing various relevant theoretical and experimental constraints, we project the surviving samples achieving the $h_2$-inflation on the planes
of $\xi_2$ versus $\lambda_2$ and $m_A-m_H$ in Fig. \ref{t1rixi2}. The non-minimal coupling parameter $\xi_2$ can be as low as 1000 for
 $\lambda_2\approx 0.26$ and a very small mass splitting between $H$ and $A$. For such case, the $\lambda_2$ sizably decreases via RG running up to the unitarity scale,
which leads to $\xi_2$ around 1000.

Now we discuss the $h_1$-inflation scenario in type-I 2HDM. Except for the inflation condition, the requirement of theory and experimental
observables for $h_1$-inflation are the same as those of $h_2$-inflation.
In Fig. \ref{t1r0xi1}, we show the surviving samples achieving the
$h_1$-inflation and satisfying various theoretical and experimental constraints. Similar to the $h_2$-inflation, the $h_1$-inflation
favors $m_A$ $(m_{H^\pm})-m_H$ to be nearly in the range of -100 GeV and 70 GeV, and $m_A$, $m_{H^\pm}$,
 $m_H$ tend to be nearly degenerate with an increase of $m_H$. The non-minimal coupling parameter $\xi_1$ can be as low as 10000
for small $\lambda_1$ and $m_A-m_H$.

\subsubsection{Higgs inflation in type-II 2HDM}
In Fig. \ref{t2rilam2} and Fig. \ref{t2ridm}, we impose the constraints of pre-inflation, and show the surviving samples achieving the $h_2$-inflation 
in the type-II 2HDM. Similar to type-I model, theoretical constraints require $\lambda_{\mid 12345 \mid}<$ 2.5, and the $h_2$-inflation
favors $\lambda_2$ to be around 0.26 for a large $\tan\beta$. $A$, $H^\pm$, and $H$ are required to have small mass splitting, 
$\mid m_A~ (m_{H\pm})-m_H \mid<$ 20 GeV, and 
$m_H$ is favored to be larger than 560 GeV since the experimental value of $b\to s\gamma$ restricts $m_{H^\pm}> 570$ GeV.
The mass splitting of $A$, $H^\pm$, and $H$ of the type-II model allowed by $h_2$-inflation is much smaller than
that of the type-I model since $m_H$ can be as low as 150 GeV in the type-I model. 

The lower-right panel of Fig. \ref{t2ridm} shows that the surviving samples satisfying the constraints at electroweak scale are located
in two different regions, i.e. the SM-like coupling region and wrong sign Yukawa coupling region. For the latter, $\sin(\beta-\alpha)$
 is larger than 0 and has a sizable deviation from 1.0. For the former, $\mid\sin(\beta-\alpha)\mid$ is very closed to 1.0.
Therefore, the factor of $s_\alpha c_\alpha$ in the wrong sign Yukawa coupling region is favored to have opposite sign from that of 
the SM-like coupling region. According to the third equation of Eq. (\ref{poten-cba}), $\lambda_3$ in the wrong sign Yukawa coupling region
can be much larger than that of the SM-like coupling region. Therefore, in the wrong sign Yukawa coupling region $\lambda_{\mid 12345 \mid}$ is much larger 
than 2.5, which breaks the theoretical constraints at high energy scale and does not achieve the $h_2$-inflation.

In Fig. \ref{t2rilhc}, we show the constraints of the direct searches for Higgs bosons at the LHC on the samples achieving the $h_2$-inflation
 in the type-II model. The $h_2$-inflation favors the surviving samples with $m_H~(m_A)>$ 560 GeV and $\mid\sin(\beta-\alpha)\approx 1.0$, 
as shown in Fig. \ref{t2ridm}. The $HVV$, $Hhh$, and $AhZ$ couplings decrease with an increase of $\mid\sin(\beta-\alpha)$.
 Therefore, the $H/A\to VV, \gamma\gamma$, $H\to hh$, and $A\to hZ$ channels do not exclude those samples. 
Similar to reason for the type-I model,
the $H/A\to t\bar{t}$, $H\to AZ$, and $A\to HZ$ channels do not impose any constraints on the parameter space.
Different from the type-I model, the down-type quark and lepton Yukawa couplings of extra Higgs bosons can be sizably
enhanced by a large $\tan\beta$. Therefore, the $b\bar{b}\to H/A \to \tau^+\tau^-$ channels can impose a lower bound on
$\tan\beta$. For example, $\tan\beta>$  10 for $m_A~(m_H)=950$ GeV.

After imposing various relevant theoretical and experimental constraints, we project the surviving samples achieving the $h_2$-inflation 
in the type-II model on the planes
of $\xi_2$ versus $\lambda_2$ and $m_A-m_H$ in Fig. \ref{t2rixi2}. Similar to type-I model, the non-minimal coupling parameter $\xi_2$ can be closed
 to 1000 for  $\lambda_2\approx 0.26$ and a very small mass splitting between $H$ and $A$.

Next we show the surviving samples achieving the
$h_1$-inflation in the type-II model and satisfying various theoretical and experimental constraints in Fig. \ref{t2r0xi1}. 
Similar to the $h_2$-inflation, the $h_1$-inflation
favors the mass splitting between $m_A$ $(m_{H^\pm})$ and $m_H$ to be approximately in the range of -20 GeV and 20 GeV, and they
 tend to be nearly degenerate with an increase of $m_H$. Similar to the type-I model, the non-minimal coupling parameter $\xi_1$ can be as low as 10000
for small $\lambda_1$ and $m_A-m_H$.

\section{Electroweak phase transition}
Now we examine the FOEWPT in the parameter space achieving Higgs inflation in type-I and type-II 2HDMs.
For the FOEWPT, the two degenerate minima will be at different points in field space and the critical temperature $T_c$, and be separated by a potential barrier.

\subsection{The thermal effective potential}
We first take $\rho_1$ and $\rho_2$ as the field configurations, and 
obtain the field dependent masses of the scalars ($h,~H,~A,~H^{\pm}$), the Goldstone boson ($G,~G^{\pm}$), the gauge boson, and fermions.
The masses of scalars are  
\begin{align}
\hm^2_{h,H} &=\rm{eigenvalues} ( \widehat{\mathcal{M}^2_P} ) \ , \\
\hm^2_{G,A} &=\rm{eigenvalues} ( \widehat{\mathcal{M}^2_A}) \ , \\
\hm^2_{G^\pm,H^\pm} &=\rm{eigenvalues}  (\widehat{\mathcal{M}^2_C})  \ ,
\end{align}
\begin{align}
\widehat{\mathcal{M}^2_P}_{11} &={3\lam_{1}\over 2} \rho^{2}_{1}+{\lam_{345} \over 2} \rho^{2}_2  + m^{2}_{12} \tb - {\lam_{1} \over 2} v^{2} \cb^2 - {\lam_{345} \over 2} v^{2} \sb^{2}\nonumber\\
\widehat{\mathcal{M}^2_P}_{22} &={3\lam_{2}\over 2} \rho^{2}_{2}+{\lam_{345} \over 2} \rho^{2}_1  + {m^{2}_{12} \over \tb} - {\lam_{2} \over 2} v^{2} \sb^2 - {\lam_{345} \over 2} v^{2} \cb^{2}\nonumber\\
\widehat{\mathcal{M}^2_P}_{12} &=\widehat{\mathcal{M}^2_P}_{21}=\lam_{345} \rho_1 \rho_2 - m^{2}_{12}\nonumber\\
\widehat{\mathcal{M}^2_A}_{11} &={\lam_{1}\over 2} \rho^{2}_{1}+ m^{2}_{12} \tb - {\lam_{1} \over 2} v^{2} \cb^2 - {\lam_{345} \over 2} v^{2} \sb^{2} +{(\lam_{3}+\lam_4-\lam_5) \over 2} \rho^{2}_2\nonumber\\
\widehat{\mathcal{M}^2_A}_{22} &={\lam_{2}\over 2} \rho^{2}_{2}+ {m^{2}_{12} \over \tb} - {\lam_{2} \over 2} v^{2} \sb^2 - {\lam_{345} \over 2} v^{2} \cb^{2} +{(\lam_{3}+\lam_4-\lam_5) \over 2} \rho^{2}_1\nonumber\\
\widehat{\mathcal{M}^2_A}_{12} &=\widehat{\mathcal{M}^2_A}_{21}=\lam_{5} \rho_1 \rho_2 - m^{2}_{12}\nonumber\\
\widehat{\mathcal{M}^2_C}_{11} &={\lam_{1}\over 2} \rho^{2}_{1}+ m^{2}_{12} \tb - {\lam_{1} \over 2} v^{2} \cb^2 - {\lam_{345} \over 2} v^{2} \sb^{2} +{\lam_{3} \over 2} \rho^{2}_2 \nonumber\\
\widehat{\mathcal{M}^2_C}_{22} &={\lam_{2}\over 2} \rho^{2}_{2}+ {m^{2}_{12} \over \tb} - {\lam_{2} \over 2} v^{2} \sb^2 - {\lam_{345} \over 2} v^{2} \cb^{2} +{\lam_{3} \over 2} \rho^{2}_1 \nonumber\\
\widehat{\mathcal{M}^2_C}_{12} &=\widehat{\mathcal{M}^2_C}_{21}={(\lam_{4}+\lam_{5}) \over 2} \rho_1 \rho_2 - m^{2}_{12},
\end{align}
where $\lambda_{345}=\lambda_3+\lambda_4+\lambda_5$.

The masses of light fermions may be safely neglected, and the masses of top quark and bottom quark are 
\begin{align}
\hm^2_t &= {1\over 2} y^2_t \rho^{2}_{2}/{s_\beta^2},  ~~\hm^2_b = {1\over 2} y^2_b \rho^{2}_{2}/{s_\beta^2}~~{\rm~for~type-I}, \nonumber\\
\hm^2_t &= {1\over 2} y^2_t \rho^{2}_{2}/{s_\beta^2},  ~~\hm^2_b = {1\over 2} y^2_b \rho^{2}_{1}/{c_\beta^2}~~{\rm~for~type-II}, 
\end{align}
with $y_t={\sqrt{2} m_t \over v}$ and $y_b={\sqrt{2} m_b \over v}.$
The masses of gauge boson are 
\begin{align}
\hm^2_{W^\pm} &= {1\over 4} g^2 \left(\rho^{2}_{1} +\rho^{2}_{2} \right),\nonumber\\
\hm^2_{Z} &= {1\over 4} (g^2+g'^2)  \left(\rho^{2}_{1} +\rho^{2}_{2} \right), \nonumber\\
\quad \hm^2_{\gamma}&=0. 
\end{align}

We take Landau gauge to calculate the thermal effective potential $V_{eff}$,
\begin{align}
V_{eff} (\rho_1,\rho_2,T)= &V_{0}(\rho_1,\rho_2) + V_{CW}(\rho_1,\rho_2) + V_{CT}(\rho_1,\rho_2) \nonumber\\
&+ V_{T}(\rho_1,\rho_2,T) + V_{ring}(\rho_1,\rho_2,T).
\label{veff0}
\end{align}
Where $V_{0}$ is the tree-level potential, $V_{CW}$ is the Coleman-Weinberg potential, 
$V_{CT}$ is the counter term, $V_{T}$ is the thermal correction, and $V_{ring}$ is the resummed daisy corrections.

The tree-level potential $V_0$ is
\begin{eqnarray} \label{v0} \mathcal{V}_{0}(\rho_1,\rho_2) &=& 
\left[{1 \over 2} m_{12}^2 t_\beta - {1 \over 4}\lam_1 v^2 c_\beta^2- {1 \over 4}\lam_{345} v^2 s_\beta^2 \right]\rho_1^2\nonumber \\
&&+ \left[{1 \over 2} m_{12}^2 {1 \over t_\beta} - {1 \over 4}\lam_2 v^2 s_\beta^2- {1 \over 4}\lam_{345} v^2 c_\beta^2 \right]\rho_2^2\nonumber \\
&&+ {\lam_1 \over 8} \rho_1^4 + {\lam_2 \over 8} \rho_2^4 - m_{12}^2 \rho_1 \rho_2 + {1\over 4} \lam_{345} \rho_1^2 \rho_2^2.
\end{eqnarray}

The Coleman-Weinberg potential in the $\overline{\rm MS}$ scheme at 1-loop level is \cite{Coleman:1973jx}:
\beq
\label{eq:CWpot}
V_{\rm CW}(\rho_{1},\rho_2) = \sum_{i} (-1)^{2s_i} n_i\frac{\hm_i^4 (\rho_{1},\rho_2)}{64\pi^2}
\left[\ln \frac{\hm_i^2 (\rho_{1},\rho_2)}{Q^2}-C_i\right],
\end{equation}
with $i=h,H,A,H^\pm,G,G^\pm,W^\pm,Z,t,b$. $s_i$ is the spin of particle i and $Q$ is a renormalization scale with $Q^2=v^2$.
The constants $C_i =\frac{3}{2}$ for scalars or fermions and
$C_i = \frac{6}{5}$ for gauge bosons.
$n_i$ is the number of degree of freedom,
\begin{align}
&n_h=n_H=n_G=n_A=1,~n_{H^\pm}=n_{G^\pm}=2,\nonumber\\
&n_{W^\pm}=6,~n_{Z}=3,~n_{t}=n_{b}=12.
\end{align}

The $V_{CW}$ term can slightly change the minimization conditions of scalar potential 
in Eq. (\ref{veff0}) and the CP-even mass matrix. To maintain the minimization conditions at T=0, we need add the counter term 
\begin{align}
\label{eq:Vct}
V_{\rm CT}(\rho_1,\rho_2)&=\delta m_1^2 \rho_{1}^2+\delta m_2^2 \rho_{2}^2 +\delta \lam_1 \rho_{1}^4+\delta \lam_{12} \rho_{1}^2 \rho_{2}^2 
+\delta \lam_2 \rho_{2}^4. 
\end{align}
The relevant coefficients are determined by
\begin{align}
\label{eq:V1der}
&\frac{\partial V_{\rm CT}}{\partial \rho_{1}} = -\frac{\partial V_{\rm CW}}{\partial \rho_{1}}\;,
 \quad \frac{\partial V_{\rm CT}}{\partial \rho_{2}} = -\frac{\partial V_{\rm CW}}{\partial \rho_{2}},\\
&\frac{\partial^{2} V_{\rm CT}}{\partial \rho_{1}\partial \rho_{1}} = - \frac{\partial^{2} V_{\rm CW}}{\partial \rho_{1}\partial \rho_{1}}\;, \quad
\frac{\partial^{2} V_{\rm CT}}{\partial \rho_{1}\partial \rho_{2}} = - \frac{\partial^{2} V_{\rm CW}}{\partial \rho_{1}\partial \rho_{2}}\;, \quad
\frac{\partial^{2} V_{\rm CT}}{\partial \rho_{2}\partial \rho_{2}} = - \frac{\partial^{2} V_{\rm CW}}{\partial \rho_{2}\partial \rho_{2}}\;, 
\end{align}
which are calculated at the electroweak minimum of $\rho_{1}=v\cb$ and $\rho_{2}=v\sb$.

It is a well-known problem that the vanishing Goldstone masses at $T=0$ in the Landau gauge will lead to
an infrared (IR) divergence due to the second derivative present in our renormalization conditions.
 To fix the divergence problem,
we take an IR cut-off at $m^2_{IR} = m^2_h$ for the Goldstone masses of the divergent terms, 
which gives a good approximation to the exact procedure of
on-shell renormalization, as discussed in \cite{PT_2HDM1.5}.

The thermal contributions $V_T$ is \cite{v1t}
\beq
\label{potVth}
 V_{\rm th}(\rho_{1},\rho_{2},T) = \frac{T^4}{2\pi^2}\, \sum_i n_i J_{B,F}\left( \frac{ \hm_i^2(\rho_{1},\rho_{2})}{T^2}\right)\;,
\eeq
where $i=h,H,A,H^\pm,G,G^\pm,W^\pm,Z,t,b$, and the functions $J_{B,F}$ are 
\beq
\label{eq:jfunc}
J_{B,F}(y) = \pm \int_0^\infty\, dx\, x^2\, \ln\left[1\mp {\rm exp}\left(-\sqrt{x^2+y}\right)\right].
\eeq

The thermal corrections with resumed ring diagrams are \cite{vdai1,vdai2}
\beq
V_{\rm ring}\left(\rho_{1},\rho_{2},T\right) =-\frac{T}{12\pi }\sum_{i} n_{i}\left[ \left( \bar{M}_{i}^{2}\left(\rho_{1},\rho_{2},T\right) \right)^{\frac{3}{2}}
-\left( \hm_{i}^{2}\left(\rho_{1},\rho_{2},T\right) \right)^{\frac{3}{2}}\right] ,
\label{eq:daisy}
\eeq
with $i=h,H,A,H^\pm,G,G^\pm,W^\pm_L,Z_L,\gamma_L$. The $W^\pm_L,~Z_L$, and $\gamma_L$ are the longitudinal gauge bosons with
$n_{W^\pm_L}=2,~n_{Z_L}=n_{\gamma_L}=1$.
The thermal Debye masses $\bar{M}_{i}^{2}\left(\rho_{1},\rho_2,T\right)$ are the eigenvalues of the full mass matrix, 
\begin{equation}
\label{eq:thermalmass}
\bar{M}_{i}^{2}\left( \rho_{1},\rho_{2},T\right) ={\rm eigenvalues} \left[\widehat{\mathcal{M}_{X}^2}\left( \rho_{1},\rho_{2}\right) +\Pi _{X}(T)\right]  ,
\end{equation}%
with $X=P,A,C$. $\Pi_X$ are  
\begin{align}
\Pi_{P11} &= \left[{9g^2\over 2} + {3g'^2\over 2} +  6\lam_{1} +4\lam_{3} +2\lam_4 \right] {T^2 \over 24}~{\rm~for~ type-I}\nonumber\\
\Pi_{P22} &= \left[{9g^2\over 2} + {3g'^2\over 2} + {6y_t^2+ 6y_b^2\over \sb^2} + 6\lam_{2} +4\lam_{3} +2\lam_4 
 \right] {T^2 \over 24}~{\rm~for~type-I}\nonumber\\
\Pi_{P11} &= \left[{9g^2\over 2} + {3g'^2\over 2} + {6y_b^2 \over \cb^2} + 6\lam_{1} +4\lam_{3} +2\lam_4 
 \right] {T^2 \over 24}~{\rm~for~ type-II}\nonumber\\
\Pi_{P22} &= \left[{9g^2\over 2} + {3g'^2\over 2} + {6y_t^2 \over \sb^2} + 6\lam_{2} +4\lam_{3} +2\lam_4 
 \right] {T^2 \over 24}~{\rm~for~type-II}\nonumber\\
\Pi_{A11} &= \Pi_{C11}=\Pi_{P11}\nonumber\\
\Pi_{A22} &= \Pi_{C22}=\Pi_{P22}\nonumber\\
\Pi_{A12} &= \Pi_{A21}=\Pi_{C12} = \Pi_{C21}=\Pi_{P12} = \Pi_{P21} =0.
\end{align}

The physical mass of the longitudinally polarized $W$ boson is 
\beq
\bar{M}_{W^{\pm}_L}^2 = {1 \over 4} g^2 (\rho^2_1+\rho^2_2) + 2 g^2 T^2.
\eeq 
The physical mass of the longitudinally polarized $Z$ and $\gamma$ boson
\beq
\bar{M}_{Z_L,\gamma_L}^2 = \frac{1}{8} (g^2+g'^2) (\rho^2_1+\rho^2_2) + (g^2 + g^{\prime 2} )T^2 \pm \Delta, 
\eeq
with 
\beq
\Delta^2 =\frac{1}{64}  (g^2 + g^{\prime 2} )^2(\rho_{1}^2 + \rho_{2}^2+8T^2)^2
- g^2 g^{\prime 2} T^2 (\rho_{1}^2 + \rho_{2}^2 + 4 T^2). 
\eeq

\subsection{Results and discussions}

\begin{figure}[tb]
\centering
 \epsfig{file=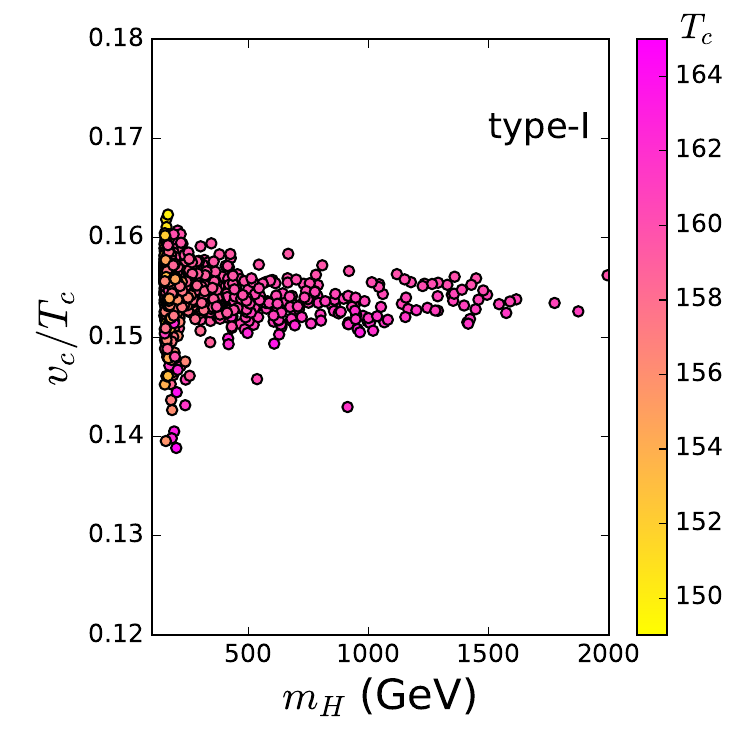,height=7cm}
 \epsfig{file=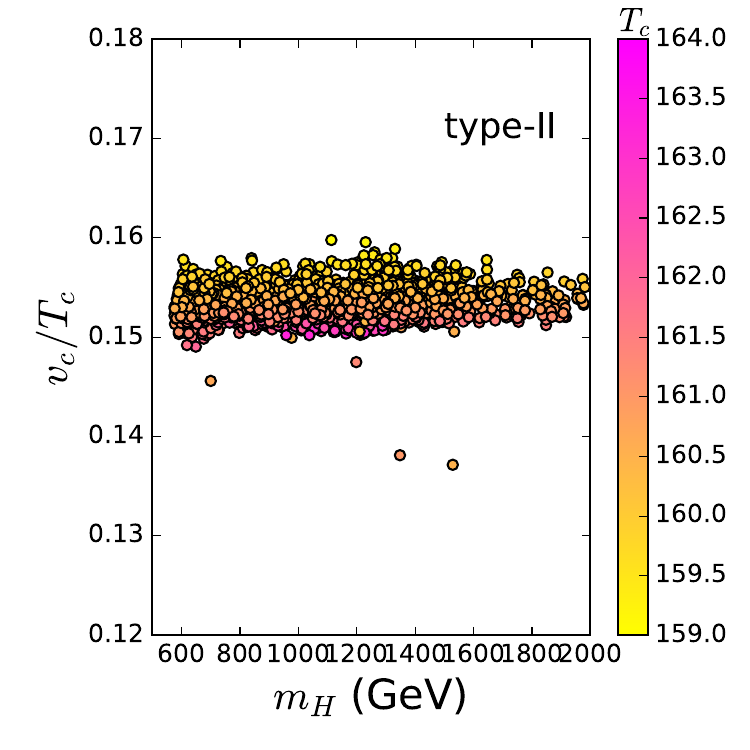,height=7cm}
\vspace{-0.5cm} \caption{In the type-I and type-II models, scatter plots of $v_c/T_c$ and $m_H$ achieving the FOEWPT, $h_2$-inflation 
or $h_1$-inflation and satisfying various relevant theoretical and experimental constraints.} \label{2hpt}
\end{figure}
The strength of EWPT is quantified as
\beq
\xi_c=\frac{v_c}{T_c},
\eeq
where $v_c=\sqrt{<\rho_1>^2+<\rho_2>^2}$ at critical temperature $T_c$. 
In order to avoid washing out the baryon number
generated during the phase transition, a strongly FOEWPT is demanded and the conventional condition is $\xi_c\geq 1$. 
We use the numerical package $\textsf{CosmoTransitions}$ \cite{cosmopt} to analyze the phase transition.

We examine whether the FOEWPT can be achieved in the parameter space of the type-I and type-II models achieving the $h_2$-inflation and $h_1$-inflation
and satisfying various relevant theoretical and experimental constraints. 
We find some surviving samples which can achieve a FOEWPT, and these samples are projected in Fig. \ref{2hpt}.
From Fig. \ref{2hpt}, we find that $\xi_c$ is always smaller than 0.17 although a FOEWPT can be achieved in the type-I and type-II models.
The thermal correction $V_T$ and the resummed daisy correction $V_{ring}$ give the cubic terms of $\phi_{1,2}$ proportional to T, 
which play key roles in generating a potential barrier and achieving the FOEWPT. Since these corrections are at one-loop level, a
strongly FOEWPT requires large $\lambda_i$. However, the Higgs inflation favors small $\lambda_i$, which leads that it is difficult to achieve the strongly FOEWPT
and Higgs inflation simultaneously.

\section{Conclusion}
We study the Higgs inflation, EWPT, and the Higgs searches at the LHC in the type-I and type-II 2HDMs with non-minimally couplings to gravity.
Imposing relevant theoretical and experimental constraints, we find that the Higgs inflation strongly restricts 
the mass splitting between $A$, $H^\pm$, and $H$, and the mass splitting tends to decrease with increasing of their masses.
In the type-I model, $m_A~(m_{H^\pm})-m_H$ is allowed to vary from -100 GeV to 70 GeV for $m_H$ around 150 GeV.
Combining the constraints of the Higgs searches at the LHC and the flavor observables, the Higgs inflation
requires $\mid m_A~(m_{H^\pm})-m_H\mid < $ 20 GeV for the type-II model.
The direct searches for Higgs at the LHC can exclude many points achieving Higgs inflation in the region of $m_H~(m_A)<$ 450 GeV in the type-I model,
and impose a lower bound on $\tan\beta$ for the type-II model. Because of the theoretical constraints, 
the Higgs inflation disfavors the wrong sign Yukawa coupling region of type-II 2HDM. 
In the region achieving the Higgs inflation, the type-I and type-II 2HDMs can achieve a FOEWPT,
but $v_c/T_c$ is much smaller than 1.0.

\section*{Acknowledgment}
We thank Bin Zhu for helpful discussions.
This work was supported by the National Natural Science Foundation
of China under grant 11975013. This work
is also supported by the Project of Shandong Province Higher Educational Science and
Technology Program under Grants No. 2019KJJ007.

\end{document}